\documentclass[12pt,a4paper]{article}
\usepackage{amsmath,amssymb,bm,ascmac,bbm,mathtools}
\usepackage[dvipdfmx, usenames]{color}
\usepackage[dvipdfmx]{graphicx}
\usepackage{xcolor}
\usepackage{here}
\usepackage{authblk}
\usepackage{lscape}
\usepackage{hyperref}

\newcommand{\tr}{\text{tr}}

\newcommand{\mcal}{\mathcal}
\newcommand{\mbb}{\mathbb}
\newcommand{\mfrak}{\mathfrak}
\newcommand{\la}{\langle}
\newcommand{\ra}{\rangle}

\newcommand{\ep}{\epsilon}

\begin{document}
\title{Symmetry enhancement in RCFT II}
\author{Ken KIKUCHI}
\affil{Yau Mathematical Sciences Center, Tsinghua University}
\date{}
\maketitle

\begin{abstract}
We explain when and why symmetries enhance in fermionic rational conformal field theories. In order to achieve the goal, we first clarify invariants under renormalization group flows. In particular, we find the Ocneanu rigidity is not enough to protect some quantities. Concretely, while (double) braidings are subject to the rigidity, they jump at conformal fixed points. The jump happens in a specific way, so the double braiding relation further constrains renormalization group flows. The new constraints enable us three things; 1) to predict infrared conformal dimensions in massless flow, 2) to reveal some structures of the theory space, and 3) to obtain a necessary condition for a flow to be massless. We also find scaling dimensions ``monotonically'' decrease along massless flows. Combining the discovery with predictions, sometimes, we can uniquely fix infrared conformal dimensions.
\end{abstract}

\tableofcontents

\makeatletter
\renewcommand{\theequation}
{\arabic{section}.\arabic{equation}}
\@addtoreset{equation}{section}
\makeatother

\section{Introduction}
Symmetry plays central roles in physics. Sometimes, symmetries are spontaneously broken (SSB) in infrared (IR). SSBs have physical consequences such as massless excitations. On the other hand, sometimes symmetries emerge in IR. Emergent symmetries are common in quantum field theory (QFT). For example, we usually define theories on a lattice in ultraviolet (UV). The systems only have discrete spacetime symmetries (translation and rotation). Then, we take its continuum (or long distance) limit expecting we get continuum QFTs. Continuum QFTs have much larger spacetime symmetries, Poincar\'e symmetry. Since the Poincar\'e symmetry is one of the axioms of QFT, understanding the mechanism behind symmetry enhancement would be one of the most fundamental problems in QFTs.

Not just spacetime symmetries, but also flavor (or internal) symmetries can enhance. One of the most famous examples is the ``$E_7$ surprise'' \cite{DG12}. Dimofte and Gaoitto found $SU(8)$ UV symmetry of certain four-dimensional supersymmetric gauge theory enhances to $E_7$ IR symmetry. As the title of the paper indicates, the enhancement was unexpected. This means we still do not understand deep structures underlying the enhancement.

In order to reveal the underlying mechanism behind the phenomena, rational conformal field theories (RCFTs) would be good places to look at. The reason is because their mathematical structures are well established; they are described by modular tensor categories (MTCs) \cite{MS1,MS2}. In fact, by limiting ourselves to renormalization group (RG) flows to RCFTs, we could explain when and why symmetries enhance \cite{K21}. We exploited constraints on RG flows from generalized global symmetries \cite{GKSW14,BT17,CLSWY}. In particular, we heavily used constraints coming from braidings.

Our proposal gave mathematically rigorous explanation of symmetry enhancement in RCFTs, however, it is not completely satisfactory; we could only explain emergent symmetries in $bosonic$ RCFTs. We also have plenty of $fermionic$ RCFTs, but our proposal cannot be immediately applied to these theories. The obstruction is in spin structures. Unlike bosonic theories, we have to keep track of spin structures in fermionic theories, and this requires some modifications of our proposal in bosonic theories.

In what follows, we thus first review spin structures on two-tori, and modular properties on them (section \ref{spinstr}). Next, we clarify (non)invariants under RG flows (section \ref{RGinv}). Along the way, we find a curious relation on double braidings; while they are ``protected'' under continuous deformations by certain discreteness, we find they jump at conformal fixed points. The jump happens in a specific way, so the double braiding relation gives us strong constraints on RG flows. We discuss three applications in section \ref{DBR}. Making the most of constrains on RG flows, we explain and test our proposal in section \ref{symenhance}. We conclude with discussion section \ref{discussion}.

We have three appendices. In Appendix \ref{symcat}, we study symmetry categories of unitary minimal models. We also discuss some RG flows employing newly obtained constraints. In Appendix \ref{TDLs}, we list topological defect lines (TDLs), i.e., generators of generalized global symmetries. We use these lines to constrain RG flows. Finally, in Appendix \ref{PMC}, we define $\mbb Z_N$ paraspin modular category, and study some of their properties.

\section{Fermionic RCFT on two-torus}\label{spinstr}
\subsection{Spin structure and modular property on tori}
Given an RCFT, one can put the theory on a two-torus $\mbb T^2$. (A useful picture for our purposes later is to view a two-torus as a parallelogram with opposite edges identified.) A two-torus admits four spin structures because $|H^1(\mbb T^2;\mbb Z_2)|=4$. Physically, this corresponds to four different choices of either anti-periodic (a.k.a. Neveu-Schwartz, or NS for short) or periodic (a.k.a. Ramond, or R for short) boundary conditions along two non-trivial cycles of $\mbb T^2$. Hence, there are $2\times2=4$ spin structures. Usually, the four spin structures are called NS-NS, NS-R, R-NS, and R-R spin structure.

If the theory is bosonic, it does not depend on spin structures. In particular, its partition function should be invariant under the whole modular group $SL(2,\mbb Z)=\la\mfrak s=\begin{pmatrix}0&-1\\1&0\end{pmatrix},\mfrak t=\begin{pmatrix}1&1\\0&1\end{pmatrix}\ra$. Historically, Cardy realized the modular invariance imposes strong constraints on operator contents of the theory \cite{C86}. On the other hand, if the theory is fermionic, it does depend on a choice of spin structure. A chosen spin structure is in general not preserved by the whole modular group, but only by its subgroup. Let us study which subgroups preserve given spin structures.

We start from NS-NS spin structure. The spin structure means anti-periodic along both non-trivial cycles of $\mbb T^2$. The modular $S$-transformation acts on the parallelogram by $\pi/2$ rotation. Thus effectively the transformation exchanges the two boundary conditions. (More precisely, two cycles $(a,b)$ are sent to $(b,-a)$. Although we can forget about the minus sign in case of $\mbb Z_2$, for larger $\mbb Z_N$ one has to be careful about the sign flip.) In the case of NS-NS spin structure, it is invariant under the exchange. Thus the modular $S$-transformation preserves the spin structure. How about the modular $T$-transformation? The transformation does not touch the boundary condition in `spatial' direction. The boundary condition in the `temporal' direction is shifted by that in `spatial' direction. In case of NS-NS spin structure, this means the modular $T$-transformation sends NS-NS spin structure to NS-R spin structure. Thus it does not preserve our NS-NS spin structure. However, if we perform the modular $T$-transformation twice, we get back the original NS-NS spin structure. Therefore, what we learned can be summarized as follows: the NS-NS spin structure is not preserved by the whole modular group $SL(2,\mbb Z)$, but only by its subgroup generated by modular $S$- and $T^2$-transformations. Such a subgroup is known as the level-two congruence subgroup\footnote{More precisely, just as $SO$ spacetime symmetry of bosonic systems are extended by fermion parity $\mbb Z_2^F=\la(-1)^F\ra$ to $Spin$ symmetry of fermionic systems, $SL(2,\mbb Z)=Sp(1,\mbb Z)$ is extended by $\mbb Z_2^F$ to the metaplectic group
\[ 1\to\mbb Z_2^F\to Mp(1,\mbb Z)\to Sp(1,\mbb Z)\to1. \]
The group is defined by the relations
\[ s^2=(st)^3,\quad s^4=f,\quad f^2=1. \]
(The second relation is easy to understand. As we explained above, modular $S$-transformation acts by $\pi/2$ rotation. Thus, $s^4$ acts by $2\pi$ rotation, and it flips signs of fermions.) However, since we only consider even spin structures, the details are not important for us. When one studies odd spin structure, i.e., R-R spin structure, one has to take the difference into account.}
\begin{equation}
    \Gamma_\theta:=\{\gamma\in SL(2,\mbb Z)|\gamma\equiv\begin{pmatrix}1&0\\0&1\end{pmatrix}\text{ or }\begin{pmatrix}0&1\\1&0\end{pmatrix}\text{ mod }2\}=\la\mfrak s,\mfrak t^2\ra\equiv\text{Cong}(\mbb T^2_\text{NS,NS}).\label{congNSNS}
\end{equation}
One important lesson is that fermionic RCFTs defined on a two-torus with NS-NS spin structure do not have well-defined topological twists $\theta_j$ (or spins $s_j$ a priori defined mod 1 because $\theta_j=e^{2\pi is_j}$) which are eigenvalues under modular $T$-transformations. On the other hand, eigenvalues under modular $T^2$-transformations should be well-defined. This means, in NS-NS spin structure, topological twists of fermionic RCFTs are defined up to sign. Equivalently, this means (topological) spins are defined mod $\frac12$, and not mod $1$. Another important fact we will use later is that the modular $S$-transformation should be invertible because it is a (projective) representation of a subgroup $\la\mfrak s\ra\subset\Gamma_\theta$.

Next, let us study R-R spin structure. Again, the spin structure is preserved by the modular $S$-transformation. Furthermore, the modular $T$-transformation also preserves the spin structure. Therefore, partition functions of fermionic RCFTs defined on a two-torus with R-R spin structure should respect the whole modular group $SL(2,\mbb Z)$ (or more precisely the metaplectic group $Mp(1,\mbb Z)$ because the spin structure is odd). Despite the extension, the key invertibility of $S$-transformations are unaffected because we still have $s^8=1$. Therefore, fermionic theories defined on the R-R spin structure should also have invertible $S$-matrices.

Finally, let us briefly see the other two spin structures, NS-R and R-NS. These spin structures are not preserved by modular $S$-transformations, rather they are exchanged; NS-R is sent to R-NS, and R-NS is sent to NS-R. This means $S^2$ preserves the spin structures. How about modular $T$-transformations? The NS-R spin structure is shifted to NS-NS, and another $T$-transformation gives us NS-R back. The R-NS spin structure is preserved by the modular $T$-transformation. To summarize, the two spin structures are not invariant under the whole modular group $SL(2,\mbb Z)$, but only under its subgroups:
\begin{equation}
\begin{split}
    \Gamma^0(2):=&\{\gamma\in SL(2,\mbb Z)|\gamma\equiv\begin{pmatrix}*&0\\ *&*\end{pmatrix}\text{ mod }2\}=\la\mfrak s^2,\mfrak t^2\ra\equiv\text{Cong}(\mbb T^2_\text{NS,R}),\\
    \Gamma_0(2):=&\{\gamma\in SL(2,\mbb Z)|\gamma\equiv\begin{pmatrix}*&*\\0&*\end{pmatrix}\text{ mod }2\}=\la\mfrak s^2,\mfrak t\ra\equiv\text{Cong}(\mbb T^2_\text{R,NS}).
\end{split}\label{congNSRRNS}
\end{equation}
These are also called the level-two congruence subgroups.

\subsection{Partition functions of fermionized RCFTs}
Facts explained in the previous subsection apply to generic fermionic RCFTs. In this subsection, we will focus on fermionic RCFTs obtained via fermionization \cite{T18} of bosonic RCFTs. This assumption practically means we start from bosonic RCFTs with anomaly-free $\mbb Z_2$ symmetry.

Let us pick a bosonic RCFT with an anomaly-free $\mbb Z_2$ symmetry generated by an invertible TDL $\eta$. The theory can be decomposed into four sectors; $\mbb Z_2$ even/odd sectors and un/twisted sectors. To see the decomposition in more detail, let us denote the partition function $Z_b$ (with subscript $b$ to remember ``bosonic''), and denote absence/presence of the $\eta$ line in time- or space-directions by 0 and 1. With these notations, the $\mbb Z_2$ even part of untwisted sector\footnote{In \cite{HNT}, the four sectors were denoted
\[ S=V^0_0,\quad T=V^1_0,\quad U=V^0_1,\quad V=V^1_1. \]
We changed the notation in order to avoid notational conflicts such as $SS,ST,TS$ where the first symbols denote modular matrices and the seconds sectors. The notation $V_j^k$ also makes clear whether it is un/twisted (recorded by the subscript $j$) or what is its charge (recorded by the superscript $k$). The notation can be readily generalized to $\mbb Z_N$.} is given by
\[ V^0_0:=\frac12(Z_b[0,0]+Z_b[0,1]). \]
The $\mbb Z_2$ odd part of untwisted sector is given by
\[ V^1_0:=\frac12(Z_b[0,0]-Z_b[0,1]). \]
Similarly, we can decompose the twisted sectors:
\[ V^0_1:=\frac12(Z_b[1,0]+Z_b[1,1]),\quad V^1_1:=\frac12(Z_b[1,0]-Z_b[1,1]). \]

With this decomposition, each partition function of the bosonic RCFT is given by
\begin{equation}
    Z_b[0,0]=V_0^0+V_0^1,\quad Z_b[0,1]=V_0^0-V_0^1,\quad Z_b[1,0]=V_1^0+V_1^1,\quad Z_b[1,1]=V_1^0-V_1^1.\label{Zbs}
\end{equation}
Let us see how each partition function transforms under modular transformations (with anomaly-free\footnote{If there exist anomalies, non-trivial phases appear \cite{NY17,CLSWY,KY19}.} $\mbb Z_2$). By definition, $Z_b[0,0]$ is invariant under both modular $S$- and $T$-transformations:
\[ SZ_b[0,0]=Z_b[0,0]=TZ_b[0,0]. \]
$Z_b[0,1]$ is invariant under $T$-transformation, while it is sent to $Z_b[1,0]$ under $S$-transformation:
\[ SZ_b[0,1]=Z_b[1,0],\quad TZ_b[0,1]=Z_b[0,1]. \]
$Z_b[1,0]$ is sent back to $Z_b[0,1]$ under $S$-transformation, and sent to $Z_b[1,1]$ under $T$-transformation:
\[ SZ_b[1,0]=Z_b[0,1],\quad TZ_b[1,0]=Z_b[1,1]. \]
$Z_b[1,1]$ is sent back to $Z_b[1,0]$ under $T$-transformation, and invariant under $S$-transformation:
\[ SZ_b[1,1]=Z_b[1,1],\quad TZ_b[1,1]=Z_b[1,0]. \]
Combining these transformation laws, we learn
\begin{equation}
    TV_0^0=V_0^0,\quad TV_0^1=V_0^1,\quad TV_1^0=V_1^0,\quad TV_1^1=-V_1^1.\label{Trans}
\end{equation}
Namely, only the $V_1^1$ sector contains spin half-integer operators. The modular $S$-transformations are more involved:
\begin{equation}
\begin{split}
    SV_0^0&=\frac12(V_0^0+V_0^1+V_1^0+V_1^1),\\
    SV_0^1&=\frac12(V_0^0+V_0^1-V_1^0-V_1^1),\\
    SV_1^0&=\frac12(V_0^0-V_0^1+V_1^0-V_1^1),\\
    SV_1^1&=\frac12(V_0^0-V_0^1-V_1^0+V_1^1).
\end{split}\label{Strans}
\end{equation}

Some fermionic RCFTs are obtained through fermionization of the bosonic RCFTs. At the level of partition function, the operation just means recombination of the four sectors. The partition functions are given by
\begin{equation}
    Z_f[0,0]=V_0^0+V_1^1,\quad Z_f[0,1]=V_0^0-V_1^1,\quad Z_f[1,0]=V_0^1+V_1^0,\quad Z_f[1,1]=-V_0^1+V_1^0,\label{Zfs}
\end{equation}
with subscript $f$ for ``fermionic.'' Their modular transformations can be easily read off from those of four sectors:
\begin{equation}
\begin{split}
    SZ_f[0,0]=Z_f[0,0],&\quad TZ_f[0,0]=Z_f[0,1],\\
    SZ_f[0,1]=Z_f[1,0],&\quad TZ_f[0,1]=Z_f[0,0],\\
    SZ_f[1,0]=Z_f[0,1],&\quad TZ_f[1,0]=Z_f[1,0],\\
    SZ_f[1,1]=&Z_f[1,1]=TZ_f[1,1]
\end{split}.\label{modularZf}
\end{equation}
One can see the partition functions have expected behaviors because
\begin{equation}
    Z_f[0,0]=Z_\text{NS-NS},\quad Z_f[0,1]=Z_\text{NS-R},\quad Z_f[1,0]=Z_\text{R-NS},\quad Z_f[1,1]=Z_\text{R-R}.\label{Zfidentify}
\end{equation}
This observation makes clear that fermionic RCFTs obtained via fermionization transform with the same modular $S$- and $T$-matrices. More precisely, characters obey the same modular transformation laws as its bosonic counterpart.\footnote{This would be the reason why \cite{BDLLS20,BDLLS21} could find characters of fermionic RCFTs from modular $S$-matrices of bosonic RCFTs.} (We will introduce different $S$-matrix later. It is important to distinguish the two.) For simplicity, we focus on NS-NS spin structure below. What is important for our purposes is the modular properties in the spin structure; 1) it should have invertible $S$-matrices; 2) (topological) spins are defined mod $\frac12$.

\section{(Non)invariants under RG flow}\label{RGinv}
In this section, we clarify what is invariant under RG flows, and what can change. We focus on braided fusion categories (BFCs). They consist of the following data:
\begin{itemize}
    \item collection of objects (i.e., TDLs in our context),
    \item fusion coefficients $N_{ij}^k$ or fusion ring,
    \item $F$-symbols,
    \item braiding or $R$-symbols.
\end{itemize}
Each item could give us (non)invariants under RG flows. We study them one by one below. Along the way, we find a curious relation between double (or full) braidings in UV and IR.

Before we get there, let us explain one thing applicable more broadly. In RG flows among QFTs without gravity, we typically do not deform background geometry. In particular, we usually do not change boundary conditions. Therefore, if we deform a fermionic theory defined on a two-torus with NS-NS spin structure, the IR theory should also be defined on the two-torus with the same spin structure. Here, one may naively think fermionic theory in UV may end up in bosonic theory in IR if all fermionic operators are lifted. However, as long as the fermion parity is preserved, one can rule out this scenario employing the spin constraint the author suggested in \cite{KCXC}. RG flows in the other spin structures can be constrained in the same way.

\subsection{Collection of objets}
Now we turn to RG (non)invariants associated to BFCs. We start from a collection of objects.

If a TDL commutes with relevant operators, it is preserved all along the RG flow \cite{G12,CLSWY}. Intuitively, this is because the line cannot `feel' whether the theory is deformed or not. The intuition should also make it clear that simplicity of lines are preserved along RG flows; a simple line cannot split along RG flows. To prove this, assume the opposite; if a simple line splits at some energy scale, then the line can `feel' the deformation, contradicting the commutativity with deformation operators. Therefore, a collection of TDLs commuting with relevant operators are preserved along the RG flow. In particular, the number of surviving lines cannot decrease, and if the collection of surviving TDLs do not satisfy some required properties (such as modularity or consistency with the $c$-theorem), the only way out is to $increase$ the number with emergent TDLs in IR. This is the key idea in \cite{K21} and in this paper.

Furthermore, we can assign numbers to TDLs by closing lines. Since the lines are topological, one can continuously shrink away the loops. After shrinking loops, one obtains quantum dimensions of the TDLs. Since the process is unaffected (for surviving TDLs) by relevant deformations, quantum dimensions are preserved all along the RG flow.

\subsection{Fusion ring}
If several TDLs commute with relevant operators, some of their properties are preserved all along the RG flow. We start with fusion rings. BFCs are equipped with fusion of objects. Namely, we can fuse two lines $\mcal L_i,\mcal L_j$. Since TDLs are topological, resulting lines $\mcal L_k$'s are also topological. The information of which lines appear with their multiplicities is encoded in the fusion coefficients $N_{ij}^k$. The coefficients (or fusion rings) are preserved along the RG flow because all lines participating the fusion commute with relevant operators.

\subsection{$F$-symbols}
Given fusions of two lines, one can fuse three lines $\mcal L_i,\mcal L_j,\mcal L_k$. Here, there are various ways to perform the fusion; either i) we first fuse $\mcal L_i,\mcal L_j$, and fuse the result with $\mcal L_k$, corresponding to $(\mcal L_i\mcal L_j)\mcal L_k$, or ii) we first fuse $\mcal L_j,\mcal L_k$, and fuse the result with $\mcal L_i$, corresponding to $\mcal L_i(\mcal L_j\mcal L_k)$. The two final results should be isomorphic. The isomorphism is given by $F$-symbols $F^{i,j,k}_{ijk}$. The $F$-symbols associated to surviving lines are also preserved along RG flows. A quickest explanation is anomaly matching; $F$-symbols encode the information of anomalies, and we know anomalies match in UV and IR. Thus, $F$-symbols should be invariant under RG flows. A more sophisticated explanation is as follows. The $F$-symbols should obey coherence relations called the pentagon axiom. It is known that solutions to the relations are discrete (up to `gauge transformations'). Therefore, the solutions are `rigid,' and they should not change under (small) continuous deformations like RG flows. This property is known as the Ocneanu rigidity \cite{ENO}. This explains why $F$-symbols are invariant under RG flows.

The $F$-symbols also fix spin contents of defect Hilbert spaces. Therefore, in an RG flow triggered by rotation-preserving relevant operators, spin contents of defect Hilbert spaces associated to surviving TDLs are also preserved (up to lifting) along the flow. More precisely, heavy operators may be lifted, so spin contents in IR should be a subset of spin contents in UV as the author found in \cite{KCXC}.

\subsection{Braidings}
Speaking of the Ocneanu rigidity, (half) braidings are also subject to the rigidity. A braiding $c$ is a natural isomorphism\footnote{Emphasizing the coherence constraints, we usually denote braidings $c_{i,j}=R_{i,j}$, and call them $R$-symbols. If $i,j$ can fuse to $k$, braiding of $i,j$ can be interpreted as fusing $i,j$ to $k$, and $k$ splitting to $j,i$. The $k$-channel is in particular denoted $R^{i,j}_k$.} $c_{i,j}:i\otimes j\to j\otimes i$. A braided fusion category (BFC) is a fusion category $\mcal C$ with a braiding $c$, $(\mcal C,c)$. The braidings are also subject to coherence relations called the hexagon axiom. It is also known that solutions to the coherence relations are discrete (up to `gauge transformations'). Therefore, one may naively expect $R$-symbols are also preserved all along RG flows. This is not true. A simple `counterexample' would be enough to understand this point. For concreteness, let us examine a known RG flow between bosonic unitary discrete series of minimal models $M(m+1,m)$. In this class of RCFTs, there is a one-to-one correspondence between TDLs (especially called Verlinde lines) and primary operators. Since the primaries are real, self-braiding (in the identity-channel) of a TDL associated to a primary $\phi_j$ with conformal dimension $h_j$ is given by topological twist
\begin{equation}
    R^{j,j}_1=\theta_j^*id_{V^{jj}_1}=e^{-2\pi ih_j}id_{V^{jj}_1}.\label{Rtheta}
\end{equation}
With this formula, let us see whether the $R$-symbols are preserved along RG flows. It is known that, with positive Lagrangian coupling, a primary operator in $M(m+1,m)$ with Kac label $(r,s)=(1,3)$, $\phi_{1,3}$, triggers RG flow to the next minimal model $M(m,m-1)$. In case of $m=4$, a TDL $N$ associated to $\phi_{2,1}$ commute with $\phi_{1,3}$, and is thus preserved along the RG flow. In the UV theory $M(5,4)$, the $N$ line has topological twist $\theta_N^\text{UV}=e^{2\pi i\frac7{16}}$, while in the IR theory $M(4,3)$, the $N$ line has topological twist $\theta_N^\text{IR}=e^{2\pi i\frac1{16}}$. Therefore, the formula (\ref{Rtheta}) tells us
\begin{equation}
    e^{-2\pi i\frac7{16}}id_{V^{NN}_1}=\left(R^\text{UV}\right)^{N,N}_1\neq\left(R^\text{IR}\right)^{N,N}_1=e^{-2\pi i\frac1{16}}id_{V^{NN}_1}.\label{RUVIR}
\end{equation}
Thus, $R$-symbols are $not$ preserved along RG flows even though they are `protected' by the Ocneanu rigidity. One way to understand the non-invariance of $R$-symbols is this; they are ``gauge dependent.'' As is evident from the identity morphism in (\ref{Rtheta}), $R$-symbols are maps acting on junction vector spaces $R^{i,j}_k:V^{ij}_k\to V^{ji}_k$, and hence depend on, say, basis of the vector spaces.

On the other hand, the double (or full) braiding $R_{j,i}R_{i,j}=c_{j,i}c_{i,j}$ is ``gauge independent.'' Therefore, one may expect the double braidings are invariant under RG flows. Is the expectation true? No. Let us re-examine the example above. The double braiding in UV is given by
\begin{equation}
    \left(R^\text{UV}\right)_{N,N}\left(R^\text{UV}\right)_{N,N}=e^{-\frac{7\pi i}4}id_1\oplus e^{\frac{5\pi i}4}id_\eta,\label{dbN}
\end{equation}
while in IR, it is given by
\[ \left(R^\text{IR}\right)_{N,N}\left(R^\text{IR}\right)_{N,N}=e^{-\frac{\pi i}4}id_1\oplus e^{\frac{3\pi i}4}id_\eta. \]
It is evident that even the double braidings are $not$ preserved along RG flows. However, since they are ``gauge independent,'' it is still meaningful to compare them. Then one immediately notices that the double braidings are the $opposite$ of each other
\[ c^\text{IR}_{N,N}c^\text{IR}_{N,N}=(c^\text{UV}_{N,N}c^\text{UV}_{N,N})^*. \]
Is this a coincidence? To make this point sure, we have computed double braidings up to $m=7$ bosonic and fermionic minimal models (in NS-NS sector). We also computed double braidings in two three-state Potts models $(D_4,A_6),(A_4,D_4)$. (For the RG flow $(D_4,A_6)\to(A_4,D_4)$, see the Appendix \ref{symcat}.) We find that double braidings of surviving lines $i,j$ are always the opposite of each other in UV and IR:
\begin{equation}
    c^\text{IR}_{j,i}c^\text{IR}_{i,j}=\left(c^\text{UV}_{j,i}c^\text{UV}_{i,j}\right)^*.\label{doublebraidIRUV}
\end{equation}
To the best of our knowledge, this relation has never appeared in literature.

Why the double braidings are the opposite in UV and IR? Or even more fundamentally, why the double braidings jump along RG flows? (When the jump happens become clear through the following consideration.) We explain these points below.

A hint is obtained by going back to our familiar example, the RG flow $M(5,4)\to M(4,3)$. While the double braiding $c_{N,N}c_{N,N}$ jumps, we know its (quantum) trace $\tr(c_{N,N}c_{N,N})$ is preserved:
\[ \tr(c^\text{IR}_{N,N}c^\text{IR}_{N,N})=0=\tr(c^\text{UV}_{N,N}c^\text{UV}_{N,N}). \]
The vanishing trace may seem special, but in fact one can show all nine traces of double braidings $\tr(c_{j,i}c_{i,j})$ for $i,j\in\{1,\eta,N\}$ are preserved along the RG flow.

An intuitive explanation special to this case (i.e., when $\mcal C$ is self-dual, meaning that $\forall j\in\mcal C,\ j^*=j$) is as follows. Recall the three-dimensional topological quantum field theory (TQFT) interpretation of two-dimensional RCFT \cite{W89}. Braiding two lines twice, one obtains locally linked strands. Note that it is not guaranteed that two lines are (globally) linked because they may be braided twice in the opposite direction far away. If that is the case, the two lines can be unlinked continuously. To really link the two lines, we have to close the ends. This is achieved by the (quantum) trace. With the (quantum) trace, two lines are closed, and they can no longer escape. Thus $\tr(c_{j,i}c_{i,j})$ gives a fundamental topological invariant of linked knots (isotopic to the Hopf link). The trace is nothing but the definition of (unnormalized) topological $S$-matrix\footnote{While this definition seems standard in math literature (e.g. \cite{M12}), it seems common in physics literature (e.g. \cite{K05}) to take dual of one object. Namely,
\[ \left(\widetilde S_\text{top}^\text{physics}\right)_{ij}:=\tr(c_{j^*,i}c_{i,j^*}). \]
Two definitions coincide when all objects are self-dual. The difference appears when there is an object $j^*\neq j$, such as the $\mbb Z_3$ objects in the three-state Potts models. We will adopt the math convention for convenience.

We would also like to emphasize that the topological $S$-matrix is in general different from the modular $S$-matrix, which describes how characters transform. We will see the difference explicitly in examples (e.g. $m=3$ fermionic minimal model) later.}
\begin{equation}
    \left(\widetilde S_\text{top}\right)_{ij}:=\tr(c_{j,i}c_{i,j}).\label{topS}
\end{equation}
This explains why the double braiding $c_{j,i}c_{i,j}$ is generically not invariant under RG flows, while its trace $\tr(c_{j,i}c_{i,j})$ is. An analogy with quantum dimensions may help a reader. When we listed invariants under RG flows, we did not talk about lines but only closed loops. The former (a line $j$) can be interpreted as $id_j$, while the latter (a closed $j$-loop) is nothing but the definition of its quantum dimension $d_j:=\tr(id_j)$. We have seen that the quantum dimensions (corresponding to closed loops) are invariant under RG flows, while we did not say lines are invariant. The difference between $\tr(c_{i,j}c_{i,j})$ vs. $c_{i,j}c_{i,j}$ is analogous to this. In fact, a quantum dimension $d_j$ is nothing but the (quantum) trace of double braiding with the trivial line $1$, $d_j=\tr(c_{j,1}c_{1,j})$. In short, for $n$ surviving TDLs in self-dual BFCs, we get $n^2$ RG invariants (not all independent) as topological $S$-matrix $\left(\widetilde S_\text{top}\right)_{ij}=\tr(c_{j,i}c_{i,j})$. This generalizes the RG invariance of quantum dimensions in the sense explained above.

At this point, a reader should ask, ``Then why $F$-symbols are invariant under RG flows without trace?'' One explanation would be as follows. In two spacetime dimensions, an $F$-move can be interpreted \cite{DW89,T17} as attaching a 3-simplex to dual triangulation, and the corresponding $F$-symbol as the Boltzmann weight assigned to the simplex. The Boltzmann weight is fixed by three TDLs associated to the edges of the simplex. What is crucial here is that the 3-simplex is closed. Hence, the Boltzmann weight does not change under continuous deformations of the simplex. To make this argument more parallel to the one in the previous paragraph, one can model the simplex by a two-sphere, which is homeomorphic to a 3-simplex. Then, one assigns the Boltzmann weight to a point wrapped by the two-sphere.\footnote{One may naively think the assignment is nothing but the $U(1)$ one-form symmetry proposed in \cite{T17}. However, this is not correct; the $U(1)$ one-form symmetry is defined in two dimensional spacetime, not three. We are trying to interpret the assignment of the Boltzmann weight to a point in three dimensions.} What is important for our argument is that the two-sphere and a point is linked in three dimensions; a point cannot escape from the two-sphere without touching it. Therefore, the Boltzmann weights, or the $F$-symbols, are topological invariants. This explains why $F$-symbols are also invariant under RG flows.

The RG invariance of topological $S$-matrix is not true in general. We find the correct relation (for surviving lines $i,j$) is
\begin{equation}
    \left(\widetilde S_\text{top}^\text{IR}\right)_{ij}=\left(\widetilde S_\text{top}^\text{UV}\right)_{ij}^*.\label{topSIRUV}
\end{equation}
This point and the jump of double braiding can be understood as follows.

Massless RG flows can be interpreted \cite{G12} as maps between observables in UV and IR CFTs. More concretely, Gaiotto proposed an interface between UV and IR CFTs. The interface is called RG interface (or RG domain wall). Using the usual folding trick, we can interpret the RG interface as RG boundary condition on $\text{CFT}_\text{UV}\times\text{CFT}_\text{IR}$. What is crucial for our purposes is that, in this process, holomorphic and anti-holomorphic sectors of one theory (say, $\text{CFT}_\text{IR}$) are exchanged. In other words, left and right are exchanged. Therefore, braiding $R_{i,j}=c_{i,j}$ defined via right-handed rule in $\text{CFT}_\text{UV}$ becomes the opposite braiding $R^{-1}_{i,j}=c^{-1}_{i,j}$ defined via left-handed rule in $\text{CFT}_\text{IR}$. This explains why double braidings are the opposite of each other in UV and IR.\footnote{This would also imply, in certain gauge, half braidings of suriving lines are the opposite in UV and IR. Indeed, we saw this in our examples above. Here, a keen reader should ask whether the opposite braiding satisfies the hexagon axiom, but it is known it does. For a given BFC $(\mcal C,c)$, $\mcal C^\text{rev}=(\mcal C,c^{-1})$ is called the reverse BFC. One may be able to draw stronger results employing the opposite half braidings. We, however, only use double braidings to give gauge-independent statements.} Since our setup --- RG flows among consecutive Virasoro minimal models --- is exactly the one considered in \cite{G12}, we believe this explains the relation (\ref{doublebraidIRUV}) for all $m$ (including exceptional models such as three-state Potts models and $E_6$ models). This picture should also make it clear when the jump happens; at the moment RG flow achieved an IR conformal point. Probably, one qualitative lesson we can learn is the following; touching conformal fixed points are not completely ``continuous.'' This also explains why emergent lines suddenly appear at the points.

This reasoning also applies to more general RG flows such as the coset theories considered in \cite{G12}. Therefore, we believe the relation (\ref{doublebraidIRUV}) also holds in those cases. It is interesting to check the relation explicitly, but we leave this point for future.

Once we have established the double braiding relation (\ref{doublebraidIRUV}), the relation between topological $S$-matrices (\ref{topSIRUV}) is an immediate consequence; just take the quantum trace of the former. The RG invariance of quantum dimensions is another consequence of this relation because, in unitary theories, we have $\left(\widetilde S_\text{top}\right)_{1j}>0$.

\section{Three applications of the double braiding relation}\label{DBR}
In this section, we would like to discuss three applications of the double braiding relation (\ref{doublebraidIRUV}). It enables us three things; 1) to predict IR conformal dimensions in massless flows, 2) to reveal some structures of the theory space, and 3) to obtain a necessary condition for a flow to be massless.

We start from predictions\footnote{Logically speaking, we have used bosonic and fermionic minimal models up to $m=7$ for check. Therefore, results beyond that should really be called predictions. In order to avoid the circular argument, we have also studied the Heptacritical Ising model $M(9,8)$. Its $\phi_{1,3}$-deformation preserves seven TDLs \cite{K21} $\{1,\mcal L_{2,1},\mcal L_{3,1},\mcal L_{4,1},\mcal L_{5,1},\mcal L_{6,1},\mcal L_{7,1}\}$. The line $\mcal L_{2,1}$ obeys fusion rule
\[ \mcal L_{2,1}\mcal L_{2,1}=1+\mcal L_{3,1}. \]
This gives us the double braiding
\[ c_{\mcal L_{2,1},\mcal L_{2,1}}c_{\mcal L_{2,1},\mcal L_{2,1}}=e^{\frac{5\pi i}8}id_1\oplus e^{-\frac{7\pi i}8}id_{\mcal L_{3,1}}. \]
If we denote the resulting IR TDLs as $\mcal L_{2,1}\to j,\mcal L_{3,1}\to k$, our double braiding relation says the corresponding TDLs should have the double braiding
\[ c_{j,j}^\text{IR}c_{j,j}^\text{IR}=e^{-\frac{5\pi i}8}id_1\oplus e^{\frac{7\pi i}8}id_k. \]
The identity-channel predicts the IR conformal dimension of the primary corresponding to the $j$ line should have
\[ e^{-4\pi ih^\text{IR}_j}=e^{-\frac{5\pi i}8}, \]
or
\[ h^\text{IR}_j=\frac5{32}\quad(\text{mod }\frac12). \]
Physically expected smallest positive choice $h_j=\frac5{32}$ is indeed satisfied by $\mcal L_{6,6}$. Similarly, the $k$-channel predicts
\[ e^{2\pi ih^\text{IR}_k}=e^{\frac{(5+7)\pi i}8}, \]
or
\[ h^\text{IR}_k=\frac34\quad(\text{mod }1). \]
Again, the prediction is satisfied with the physically expected smallest positive one $h_k=\frac34$ by $\mcal L_{6,5}$.} on conformal dimensions $h^\text{IR}$'s of IR primaries corresponding to surviving TDLs. To understand this point, let us again consider our familiar example, RG flow from tricritical Ising model triggered by $\phi_{1,3}$. We saw the $N$ line is preserved. Now let us pretend we do not know conformal dimensions in IR. In UV, the double braiding was given by (\ref{dbN}). The identity-channel $e^{-\frac{7\pi i}4}id_1$ tells us that, in IR, the $N$ line corresponds to a primary operator with topological twist
\[ \theta_N^{-2}=e^{-4\pi ih_N^\text{IR}}=e^{-\frac{\pi i}4}. \]
Solving this for $h_N^\text{IR}$, we obtain\footnote{Note that if one naively equates double braidings in UV and IR believing that the Ocneanu rigidity always gives RG invariants, the prediction on IR conformal dimension becomes $h_N^\text{IR}=-\frac1{16}$ mod $\frac12$. Assuming the unitariry, we get $h_N^\text{IR}=\frac7{16},\frac{15}{16},\frac{23}{16},\dots\ .$ We know this is wrong.}
\[ h_N^\text{IR}=\frac1{16}\quad(\text{mod }\frac12). \]
With the assumption of unitarity, this means
\[ h_N^\text{IR}=\frac1{16},\frac9{16},\frac{17}{16},\dots\ . \]
Furthermore, remember that conformal dimensions enter energy $L_0|h\ra=h|h\ra$. Therefore, it is physically expected that the smallest positive candidate is chosen. Then, we get a prediction $h_N^\text{IR}=\frac1{16}$ as advertised. Indeed, we know this conformal dimension is realized by the spin operator in the critical Ising model.

Similarly, let us next predict the IR conformal dimension corresponding to the $\eta$ line; the $\eta$-channel tells us that we must have
\[ \theta_\eta=e^{2\pi ih_\eta^\text{IR}}=e^{\frac{(1+3)\pi i}4}, \]
or\[ h_\eta^\text{IR}=\frac12\quad(\text{mod }1). \]
We achieved the stronger predictive power (i.e., mod $1$ and not mod $\frac12$) because the topological twist $\theta_\eta$ enters the double braiding linearly. The principles of unitarity and the lowest energy suggest $h_\eta^\text{IR}=\frac12$. We again get the correct conformal dimension realized by the energy operator in the critical Ising model. Since conformal dimensions are fundamental conformal data characterizing CFTs, our predictions strongly constrain possible IR theories.

Before we move on to the second application of the double braiding relation, let us further mention an observation on ``monotonicity'' of scaling dimensions.\footnote{There is a similar statement in conformal quantum mechanics \cite{O17}, which is closer to the $c$-theorem, but we are not aware of literature which discussed ``monotonicity'' of scaling dimensions in QFTs.} In Wilsonian RG \cite{WK73}, we introduce a UV cutoff $\Lambda_\text{UV}$, and integrate out modes with momenta just below $\Lambda_\text{UV}$. As a result, low energy observers have access to lower and lower energies as we follow RG flows to IR. Therefore, we intuitively expect scaling dimensions of observables in effective theories would decrease monotonically. In fact, in special class of RG flows, we can prove this intuition. Let us consider the whole sequence of RG flows among bosonic unitary discrete series of minimal models $M(m+1,m)$. In \cite{K21}, we proved that exactly $(m-1)$ lines in $M(m+1,m)$ with Kac indices $(r,1)$ for $r=1,2,\dots,m-1$ survive the relevant $\phi_{1,3}$-deformation. We also matched the surviving lines in UV $M(m+1,m)$ and IR $M(m,m-1)$:
\begin{equation}
    \begin{array}{cccccc}
    \text{UV}:&\mcal L_{1,1}&\mcal L_{2,1}&\cdots&\mcal L_{m-2,1}&\mcal L_{m-1,1}\\
    &\downarrow&\downarrow&\cdots&\downarrow&\downarrow\\
    \text{IR}:&\mcal L_{1,1}&\mcal L_{m-2,m-2}&\cdots&\mcal L_{m-2,2}&\mcal L_{m-2,1}
    \end{array}.\label{matchingTDLs}
\end{equation}
The primary corresponding to the identity line $1$ has conformal dimension $h_1=0$. Thus, we trivially have $h^\text{UV}_1=h^\text{IR}_1$. What is more interesting is non-trivial lines. For $\mcal L_{r,1}$ with $r=2,\dots,m-1$ in UV, the corresponding primary has conformal dimension $h^\text{UV}_{r,1}=\frac{[(m+1)r-m]^2-1}{4m(m+1)}$. The matching (\ref{matchingTDLs}) claims the line flows to $\mcal L_{m-2,m-r}$ associated to the primary with conformal dimension $h^\text{IR}_{m-2,m-r}=\frac{[m(m-2)-(m-1)(m-r)]^2-1}{4m(m-1)}$. Their difference is given by
\begin{equation}
    h^\text{UV}_{r,1}-h^\text{IR}_{m-2,m-r}=\frac{r^2-1}{2m}.\label{hUV-hIR}
\end{equation}
This is positive for $r>1$. We proved the ``monotonicity''\footnote{Rigorously speaking, we should call the property monotonic if we could show derivatives of scaling dimensions with respect to RG scale are non-negative. However, since we are only interested in conformal points, we abuse the word (with quotation marks) for simplicity.}
\begin{equation}
    h^\text{UV}-h^\text{IR}\ge0\label{hUV>hIR}
\end{equation}
of scaling dimensions under RG flows for the whole sequence. It would be interesting to generalize this observation to generic RG flows. We leave this point for future.

Combining the ``monotonicity'' (\ref{hUV>hIR}) and the double braiding relation, we can often fix IR conformal dimensions uniquely.\footnote{If one finds no suitable candidate, one can rule out massless phases in this way. For example, let us consider the deformation problem of critical three-state Potts model $(A_4,D_4)$. The primary $Z_1$ (or $Z_2$) corresponding to the $\mbb Z_3$ line $\eta$ breaks $\eta$ while it preserves the Fibonacci line $W$. The $W$ line corresponds to $\ep$ with conformal dimension $\frac25$. The double braiding of $W$ with itself is given by
\[ c_{W,W}^\text{UV}c_{W,W}^\text{UV}=e^{\frac{2\pi i}5}id_1\oplus e^{-\frac{4\pi i}5}id_W. \]
Now, let us assume the theory flows to a CFT. Denoting the IR TDL as $W\to j$, the line must have a double braiding
\[ c_{j,j}^\text{IR}c_{j,j}^\text{IR}=e^{\frac{-2\pi i}5}id_1\oplus e^{\frac{4\pi i}5}id_j. \]
The $j$-channel gives prediction on IR conformal dimension $h^\text{IR}_j$: 
\[ \theta_j^{-1}=e^{-2\pi ih^\text{IR}_j}=e^{\frac{4\pi i}5}, \]
or
\[ h^\text{IR}_j=\frac35\quad(\text{mod }1). \]
Here, the ``monotonicity'' claims $h^\text{IR}_j\le\frac25$. Logical possibilities $h^\text{IR}_j=-\frac25,-\frac75,-\frac{12}5,\dots$ are forbidden in unitary theories. We conclude the flow is massive. Note that we cannot draw this result from the reality condition discussed below because the $W$ line is emergent (see the Appendix \ref{symcat}).} For example, let us again look at our friend, $M(5,4)\to M(4,3)$. Our prediction on $h^\text{IR}_N$ combined with $h^\text{IR}_N\le\frac7{16}$ uniquely gives $h^\text{IR}_N=\frac1{16}$. Similarly, we can fix $h^\text{IR}_\eta=\frac12$ by combining $h^\text{IR}_\eta=\frac12$ mod $1$ with $h^\text{IR}_\eta\le\frac32$. (Since non-trivial primaries are generally renormalized, the equality would be satisfied only by special operators such as the identity.)

Here is one technical remark. We achieved predictions on IR conformal dimensions mod $\frac12$ or mod $1$. In bosonic theories where (topological) spins are defined mod $1$, we can use the full predictive power. However, if (topological) spins are defined mod $\frac12$ as in fermionic theories defined on a two-torus with NS-NS spin structure, it is meaningless to talk about (topological) spins mod $1$. Rather, the best we can achieve is prediction mod $\frac12$. Since our main focus is the latter case, our predictions are given mod $\frac12$ in the body of the paper.

The second application of the double braiding relation (\ref{doublebraidIRUV}) is as follows; TDLs can survive sequence of RG flows only if they have real double braidings. To show this, let us pick three RCFTs, $\text{RCFT}_1,\text{RCFT}_2,\text{RCFT}_3$ with RG flows $\text{RCFT}_1\to\text{RCFT}_2$ and $\text{RCFT}_2\to\text{RCFT}_3$. We can view the sequence of RG flows $\text{RCFT}_1\to\text{RCFT}_2\to\text{RCFT}_3$ as a single RG flow $\text{RCFT}_1\to\text{RCFT}_3$.\footnote{Formally, this would be realized as a composition of two maps given by RG interfaces.}
If the relation (\ref{doublebraidIRUV}) applies to these three RG flows, lines $i,j$ surviving the sequence of RG flows should obey
\[ \left(c_{j,i}^{\text{RCFT}_1}c_{i,j}^{\text{RCFT}_1}\right)=\left(c_{j,i}^{\text{RCFT}_2}c_{i,j}^{\text{RCFT}_2}\right)^*=\left(c_{j,i}^{\text{RCFT}_3}c_{i,j}^{\text{RCFT}_3}\right)=\left(c_{j,i}^{\text{RCFT}_3}c_{i,j}^{\text{RCFT}_3}\right)^*, \]
where each equality is coming from the corresponding RG flow. From the last equality, we see the double braiding has to be real
\begin{equation}
    \left(c_{j,i}c_{i,j}\right)=\left(c_{j,i}c_{i,j}\right)^*\label{realcc}
\end{equation}
for lines $i,j$ surviving the sequence of RG flows. The identity line $1$ and a $\mbb Z_2$ line $\eta$ satisfy the conditions. This would explain why only the two lines in unitary discrete serices of bosonic minimal models $M(m+1,m)$ survive sequence of RG flows down to $m=3$. (Non-invertible lines typically have non-real double braidings.) In other words, the condition explains why the flows are ``connected.'' Simultaneously, the reality condition also explains why three-state Potts models are ``disconnected.'' Since the RG flow between them preserves $\mbb Z_3$ lines with non-real double braidings, the flow should terminate at the lower CFT. More generally, $\mbb Z_N$-symmetric RG flows with non-real monodromy charges (see the appendix \ref{PMC}) should terminate after one flow.\footnote{Some theories have $\mbb Z_N$ objects with real double braidings. For example, the $(E_6)_3$ Wess-Zumino-Witten (WZW) model has $\mbb Z_3$ (center) symmetry. Its generators labeled with $3\hat\omega_0,3\hat\omega_1,3\hat\omega_5$ have integer conformal dimensions ($0,2,2$, respectively), and hence the rank three $\mbb Z_3$ BFC has double braidings
\[ \begin{pmatrix}id_1&id_\eta&id_{\eta^2}\\id_\eta&id_{\eta^2}&id_1\\id_{\eta^2}&id_1&id_\eta\end{pmatrix}. \]
If there exists $\mbb Z_3$ symmetric massless RG flow, it does not have to terminate.}

This observation gives the third application of the double braiding relation; constraints on IR phases. More precisely, the constraint gives a necessary condition to achieve massless phases. This works when we look at a sequence of RG flows as follows. Let us pick an RCFT, $\text{RCFT}_1$, and consider a massless RG flow to a lower RCFT, $\text{RCFT}_2$. We write the surviving BFC $\mcal C$. Now suppose there exists a pair of lines $i,j\in\mcal C$ with non-real double braiding. ($i$ and $j$ can be the same.) This means $c_{j,i}c_{i,j}\neq(c_{j,i}c_{i,j})^*$. Then, the pair $i,j\in\mcal C$ cannot survive the next massless RG flow to even lower RCFT, $\text{RCFT}_3$. This is because otherwise it contradicts the reality. Therefore, assuming the double braiding relation holds, we learn relevant deformations of the $\text{RCFT}_2$ preserving the pair $i,j\in\mcal C$ inherited from the upper RG flow $\text{RCFT}_1\to\text{RCFT}_2$ should end up in gapped phases. In other words, to achieve massless RG flows, relevant deformations should be chosen so that all pairs of non-emergent lines with non-real double braidings (which is typical for non-invertible lines) are broken. On the other hand, emergent lines are not constrained by upper RG flows.\footnote{This is similar to 't Hooft anomalies in emergent symmetries. An example is the flow from the fermionic $m=5$ to $m=4$ minimal model. The emergent $R$ and $(-1)^FR$ lines have 't Hooft anomalies.} Indeed, this is the case for the whole sequence of RG flows (\ref{matchingTDLs}). Since only two lines at two ends are invertible, $(m-3)$ lines in the middle are non-invertible. Here, we can prove they have non-real double braidings.\footnote{A proof is as follows. All the non-invertible lines $\mcal L_{r,1}$ with $r=2,3,\dots,m-2$ have identity-channels. Thus it is enough if we could show $\theta_{\mcal L_{r,1}}^{-2}$ is non-real. The lines have topological twists $\theta_{\mcal L_{r,1}}=e^{2\pi ih_{r,1}}$ with $h_{r,1}=\frac{[(m+1)r-m]^2-1}{4m(m+1)}$. This means the identity-channel is real iff $\frac{[(m+1)r-m]^2-1}{m(m+1)}\in\mbb Z$. We denote $n:=\frac{[(m+1)r-m]^2-1}{m(m+1)}$. Assuming $n$ is an integer, we can show a contradiction. First, the denominator of $n$ is positive. Furthermore, since $r\neq1$, the numerator is also positive, and we must have $n>0$. On the other hand, we need $1-m<n<-1$ as follows. Let us view the definition of $n$ as a quadratic equation of $r$, and solve for it. Then we get
\[ r=m\pm\sqrt{m^2+m(n-1)+1}. \]
For $r$ to be in $\{2,3,\dots,m-2\}$, we have to choose the minus sign. In addition, for $r$ to be an integer, we need $m^2+m(n-1)+1=N^2$ for some integer $N$. Then we have $r=m-|N|$. The integer should be in the range $1<N^2<(m-1)^2$ for $r$ to be in the range. Now view this as inequalities on $n$: $1<m^2+m(n-1)+1<(m-1)^2$. We get $1-m<n<-1$ in contradiction with $n>0$. Thus, such $n$ does not exist, and the identity-channel is non-real. This implies surviving non-invertible lines always have non-real double braidings.} Therefore, from the reality condition, they cannot be preserved in the next massless flow, and indeed one notices that, for $m>3$,\footnote{For $m=3$, there are no lower unitary CFT, and it is known that $\phi_{1,3}$-deformation is massive.} all of them are broken in the next massless flow. At the same time, this constraint on RG flows gives another explanation when and why emergent lines should appear; if there exists an upper RG flow $\text{RCFT}_1\to\text{RCFT}_2$ to our UV theory $\text{RCFT}_2$, lines with non-real double braidings surviving the next massless RG flow $\text{RCFT}_2\to\text{RCFT}_3$ should be emergent in the upper RG flow. In case of the sequence (\ref{matchingTDLs}), this requires $\mcal L_{2,1},\dots,\mcal L_{m-2,1}$ be emergent. We know this is true.

As opposed to massless RG flows, we can also expose our claim to tests in massive RG flows. Our necessary condition claims an RG flow should be massive if it preserves non-emergent lines with non-real double braidings. In fact, it is known that $\sigma'$ of tricritical Ising model $M(5,4)$ triggers massive RG flows (with two vacua). We can now understand why the flow is not massless; the relevant operator preserves the Fibonacci line $W$, which is non-emergent and has non-real double braiding with itself, $c_{W,W}c_{W,W}\neq(c_{W,W}c_{W,W})^*$. This contradicts the reality condition if the flow were massless. We can give another evidence of our claim. There is a numerical method to study IR phases of deformed CFTs. One of the most famous methods is called the truncated conformal space approach \cite{YZ89}. The author wrote a code for \cite{KCXC} based on the STRIP \cite{LM91}. The code can be readily applied to deformation problems of bosonic minimal models. Let us consider the tetracritical Ising model $M(6,5)$. The numerical results suggest $\ep'$ triggers massive RG flow with three and two vacua for positive and negative Lagrangian couplings, respectively. The relevant operator $\ep'$ preserves the non-invertible line $M$, which is non-emergent and has non-real double braiding with itself. Therefore, the gapped phase can be understood as a result of the reality constraint.

Other consequences of the double braiding relation will be reported elsewhere.

\section{Symmetry enhancement in fermionic RCFT}\label{symenhance}
\subsection{Proposal}
After we have clarified (non)invariants under RG flows, we can make the most of them to constrain RG flows. In particular, it is fairly simple to state when and why symmetries should enhance in fermionic RCFTs. For simplicity, we limit ourselves to the NS-NS spin structure, but generalization to the other spin structures is straightforward; just modify the congruence subgroup appropriately. Then, in NS-NS spin structure, symmetry category of IR RCFT\footnote{Note that here we are $assuming$ rationality. This assumption does not hold in general, but fortunately the assumption is satisfied in our examples because all lower CFTs are rational.} should be a $\Gamma_\theta$-BFC\footnote{We will call a BFC with topological twists defined up to sign and with invertible topological $S$-matrix as $\Gamma_\theta$-BFC because it is cumbersome to repeat the long adjectives. We looked for a name for such a BFC, but unfortunately we could not find one in literature. When one is interested in the other spin structures, one just replaces $\Gamma_\theta$ with an appropriate congruence subgroup.} with $c<c_\text{UV}$. The key invertibility of topological $S$-matrices is a consequence of the fact that the assignments
\[ \mfrak s\mapsto\widetilde S_\text{top},\quad\mfrak t^2\mapsto T^2 \]
define a projective representation of the level-two congruence subgroup $\Gamma_\theta$. (See \cite{fMTC}.) In particular, since $\mfrak s$ is invertible, $\mfrak s^4=1$, topological $S$-matrices should also be. If surviving BFCs have either non-invertible topological $S$-matrix, or are inconsistent with the $c$-theorem, the symmetry category should be enlarged so that the two conditions are satisfied simultaneously. We will test our proposal below.

To perform the tests, it is crucial to judge whether a surviving BFC has invertible topological $S$-matrix or not. Here, we employ the Lemma E.13 of \cite{K05}; a line $i$ is transparent in $\mcal C$ iff for all $j\in\mcal C$, $\frac{\left(S_\text{top}\right)_{ji}}{\left(S_\text{top}\right)_{1i}}=d_j\equiv\frac{\left(S_\text{top}\right)_{j1}}{\left(S_\text{top}\right)_{11}}$ where $S_\text{top}$ is the normalized topological $S$-matrix $S_\text{top}:=\widetilde S_\text{top}/D$ with $D^2:=\sum_jd_j^2$. Since the quantum dimensions are nonzero, we can rewrite it as
\begin{equation}
    i\in\mcal C'\equiv Z_2(\mcal C)\iff\forall j\in\mcal C,\ M_{ij}:=\frac{\left(S_\text{top}\right)_{ij}\left(S_\text{top}\right)_{11}}{\left(S_\text{top}\right)_{1i}\left(S_\text{top}\right)_{1j}}=1.\label{M}
\end{equation}
If $i$ or $j$ is invertible (tradionally called simple currents in CFT literature), one can prove $M_{ij}$ is a phase. In these cases, $M_{ij}$ was called monodromy charge \cite{SY90,FRS04,fMTC}. Invertibility of one of the fusing lines $i,j$ guarantees simplicity of the resulting line $ij$. Then one can use a convenient formula in \cite{fMTC}. However, the criterion (\ref{M}) is valid even when both $i$ and $j$ are non-invertible. We will abuse the terminology, and call $M$ monodromy charge matrix. The equation (\ref{M}) gives a useful criterion to judge invertibility of the topological $S$-matrices.

Our criterion (\ref{M}) may seem mathematical. However, we can give a physical interpretation. Note that the monodromy charge matrix $M_{ij}$ can be written as
\[ M_{ij}=\frac{\left(S_\text{top}\right)_{ij}/\left(S_\text{top}\right)_{11}}{\Big(\left(S_\text{top}\right)_{1i}/\left(S_\text{top}\right)_{11}\Big)\Big(\left(S_\text{top}\right)_{1j}/\left(S_\text{top}\right)_{11}\Big)}=\frac{\left(\widetilde S_\text{top}\right)_{ij}}{d_id_j}. \]
(This is the formula given in \cite{fMTC}.) Recall that the topological $S$-matrix is nothing but the expectation value of two linked knots. A loop of $i$ is transparent iff loops of $i$ and $j$ can be unlinked freely for any $j$. Then one can shrink away the loops $i,j$ to get (a product of) quantum dimensions $d_id_j$. The product is canceled by the same product in the denominator to give $1$. Therefore, $M_{ij}=1$ for any $j$ simply means $i$ loops can be freely unlinked.

\subsection{Test: NS-NS spin structure}
Let us expose our proposal to tests. Here, we study fermionic minimal models in NS-NS spin structure. In particular, we will focus on their relevant deformations with $|\phi_{1,3}|^2$. We choose the specific relevant operator because we can argue the IR theory would be an RCFT (for one sign of relevant coupling).

One argument employs duality. Two theories $T_1,T_2$ are said to be dual if they flow to the same IR theory $T_3$. In particular, $T_1$ (or $T_2$) and $T_3$ are also tautologically dual. In other words, theories $T_1,T_2$ belong to the same universality class represented by $T_3$, in which $T_3$ itself belongs. Furthermore, it is expected that if we perform the same manipulations to dual theories, they lead to new (possibly the same) dual theories. (See, e.g. beautiful lectures by Seiberg \cite{S16}.) Now let us come back to our problem. In bosonic theories, we know $\phi_{1,3}$-deformation of $M(m+1,m)$ leads to $M(m,m-1)$. In the terminology above, $M(m+1,m)$ and $M(m,m-1)$ are dual.\footnote{When we speak of duality, we have to specify maps between operators. Luckily, such maps were studied in detail in \cite{G12}.} Then, we perform fermionization to the two theories to get fermionic $m$ and $(m-1)$ minimal models. They are again expected to be dual. This means the higher theory flows to the lower theory. We gave analytical and numerical evidences of the flow in \cite{KCXC} for $m=4,5$.

Another argument (which would hopefully be more convincing) employs commutativity of relevant deformation and discrete gauging. Gauging a global $\mbb Z_2$ symmetry\footnote{The discussion also works for $\mbb Z_N$.} is performed as follows; insert $\mbb Z_2$ generating line $\eta$'s, and sum up the resulting partition functions. If relevant operators commute with the line $\eta$, then two results should match regardless one first inserts $\eta$ lines and deforms later, or deforms first and inserts $\eta$ lines later. To be more concrete, let us study a theory $T$ with twisted partition function $Z_T[j,k]$ in UV. If $\eta$ commutes with the relevant operators, we should have
\[ \left(Z_T[j,k]\right)'=\left(Z_T\right)'[j,k], \]
where the prime $'$ means the relevant deformation. Namely, we can change the order of relevant deformations and insertions of $\eta$ lines. This applies to all twisted partition functions. Thus, we end up having
\begin{equation}
\begin{split}
    \left(Z_{T/\mbb Z_2}\right)'&\equiv\frac12\left(Z_T[0,0]+Z_T[0,1]+Z_T[1,0]+Z_T[1,1]\right)'\\
    &=\frac12\left\{\left(Z_T\right)'[0,0]+\left(Z_T\right)'[0,1]+\left(Z_T\right)'[1,0]+\left(Z_T\right)'[1,1]\right\}\equiv\left(Z_T\right)'_{/\mbb Z_2}.
\end{split}\label{commute}
\end{equation}
Recently, this argument was extensively used in \cite{CCHLS}.

Since fermionization is a kind of discrete gauging (more precisely, one first multiplies the Arf theory, and gauges the diagonal $\mbb Z_2$), it is expected that $\phi_{1,3}$-deformation of the fermionic minimal model leads to the next fermionic minimal model (with one sign of the relevant coupling). More precisely, at the level of partition function, fermionization is achieved by flipping one sign of the summand. (We flip the sign of the fourth twisted partition function to get fermionic theory on NS-NS spin structure.) Since all twisted partition functions commute with relevant deformation separately, the equality should hold even when one changes some signs.

To summarize, we have various arguments and evidences that $\phi_{1,3}$-deformation (for one sign of the relevant coupling) of the fermionic $m$ minimal model leads to the next $(m-1)$ theory. We will use these flows to check our proposal.

The first two examples with $m=4,5$ were studied in \cite{KCXC}. For larger $m$, there are many lines, and we find it convenient to label them with conformal dimensions (except $1$, $(-1)^F$, and $\mcal C$). We hope this can avoid possible notational confusions (such as possible exchange of Kac indices $r\leftrightarrow s$). We also label primaries with their conformal dimensions.

Tests proceed as follows. We first list TDLs (see the appendix \ref{TDLs}). In this process, we also study their actions on primaries. Using the action, we read off which lines are preserved under the relevant $|\phi_{1,3}|^2$-deformation. The surviving TDLs constrain the RG flow. Since our main focus is symmetry enhancement, we are especially interested in the monodromy charge matrices. We find the TDL generating fermion parity $(-1)^F$ is transparent when $m$ is odd. (Unlike bosonic cases \cite{K21}, we could not prove this because we do not have generic formula for $S$-matrices. It would be useful to find the formula, and prove this observation. We leave these points for future.) When $m$ is even, the surviving TDLs have invertible topological $S$-matrices. Thus we cannot say symmetries should enhance just from the monodromy charge matrices. Here, in bosonic cases, we studied central charges associated to surviving MTCs. However, in fermionic theories, $\Gamma_\theta$-BFCs are poorly classified, and in particular, their central charges are unavailable. If we were unaware of the double braiding relation, we had to stop here. Fortunately, however, we realized stronger constraints from double braiding relation,\footnote{Here is one caveat. As we explained above, in NS-NS spin structure, topological twists are defined up to sign. Thus double braidings (and hence topological $S$-matrix elements) suffer from sign ambiguity. To deal with this problem, we first notice that signs of double braidings with the trivial line $1$ can be fixed thanks to the unitarity $\left(\widetilde S_\text{top}\right)_{1j}\equiv\tr(c_{j,1}c_{1,j})=d_j>0$. Then we can fix signs of the other elements using the action of line $k$ on another line $j$, $\frac{\left(\widetilde S_\text{top}\right)_{kj}}{\left(\widetilde S_\text{top}\right)_{1j}}$. We do not have a proof that this always works, but the method works in examples we study below. Related to this, identifications of TDLs suffer from ambiguities. As a working hypothesis, we label TDLs borrowing knowledge from bosonic theories. Regardless of the ambiguities, sets of constraints on RG flows are unaffected.} and indeed they save us; we can explain symmetry enhancement without resorting to central charges. Let us see this in concrete examples with small $m$. Since minimal models have ``mod 4 structure,'' it is enough to study four examples.
\begin{itemize}
    \item $m=4$:\\
    The theory has eight TDLs generated by three primitive lines $\{(-1)^F,W,R\}$. The deformation operator $|\phi_{1,3}|^2=|\ep_{\frac35}|^2$ commutes with four TDLs $\{1,(-1)^F,R,(-1)^FR\}$. Let us first check whether the IR theory can be gapless. One immediately realizes the massless scenario is allowed because all lines are invertible, and in particular the rank four surviving BFC has real double braidings. Hence, $|\phi_{1,3}|^2$ can trigger massless RG flow. The $c$-theorem tells us that the only candidates are fermionic and bosonic $m=3$ minimal models with $c=\frac12$. We can rule out the bosonic theory because it does not have four simple lines. Therefore, the only possible IR CFT is the $m=3$ fermionic minimal model.
    
    Now, let us come to our main point, symmetry enhancement. The surviving lines have monodromy charge matrix
    \begin{equation}
        M=\begin{pmatrix}1&1&1&1\\1&1&-1&-1\\1&-1&-1&1\\1&-1&1&-1\end{pmatrix}.\label{M4}
    \end{equation}
    One sees there is no nontrivial transparent line. Therefore, the surviving rank four BFC has invertible topological $S$-matrix, and our proposal claims symmetry enhancement is unnecessary unless the category cannot have central charge smaller than $c_\text{UV}=\frac7{10}$. Unfortunately, it seems central charge of the rank four $\Gamma_\theta$-BFC is unavailable in literature. However, we notice that (\ref{M4}) is nothing but the unnormalized topological $S$-matrix of the fermionic $m=3$ minimal model (in the basis $\{1,(-1)^F,(-1)^{F_R},(-1)^{F_L}\}$)
    \[ \widetilde S_\text{top}^{m=3}=\begin{pmatrix}1&1&1&1\\1&1&-1&-1\\1&-1&-1&1\\1&-1&1&-1\end{pmatrix} \]
    because all lines have quantum dimensions one. Therefore, we believe the rank four $\Gamma_\theta$-BFC supports central charge $c=\frac12$. If that is the case, symmetry enhancement is unnecessary consistent with the known RG flow \cite{KCXC}. This example should also make it clear that modular and topological $S$-matrices are different in general.
    
    The spin constraint gives the identifications
    \begin{equation}
    \begin{array}{ccccc}
    \text{UV}:&1&(-1)^F&R&(-1)^FR\\
    &\downarrow&\downarrow&\downarrow&\downarrow\\
    \text{IR}:&1&(-1)^F&(-1)^{F_R}&(-1)^{F_L}
    \end{array}.\label{matchingm=4}
    \end{equation}
    \item $m=5$:\\
    The theory has 10 TDLs generated by four primitive lines $\{(-1)^F,M,W,N\}$. The deformation operator $|\phi_{1,3}|^2=|\ep_{\frac23}|^2$ commutes with four TDLs $\{1,(-1)^F,W,(-1)^FW\}$. In \cite{KCXC}, we showed positive Lagrangian coupling triggers massless RG flow to the $m=4$ fermionic minimal model with a help of numerical method. However, since non-invertible lines are preserved, we can make analytic constraints more severe. In particular, we can predict IR conformal dimensions employing our double braiding relation. The double braidings are given by
    \begin{equation}
       \begin{pmatrix}id_1&id_{(-1)^F}&id_W&id_{(-1)^FW}\\id_{(-1)^F}&id_1&id_{(-1)^FW}&id_W\\id_W&id_{(-1)^FW}&e^{\frac{2\pi i}5}id_1\oplus e^{-\frac{4\pi i}5}id_W&e^{\frac{2\pi i}5}id_{(-1)^F}\oplus e^{-\frac{4\pi i}5}id_{(-1)^FW}\\id_{(-1)^FW}&id_W&e^{\frac{2\pi i}5}id_{(-1)^F}\oplus e^{-\frac{4\pi i}5}id_{(-1)^FW}&e^{\frac{2\pi i}5}id_1\oplus e^{-\frac{4\pi i}5}id_W\end{pmatrix}.\label{DB5}
    \end{equation}
    The double braiding relation (\ref{doublebraidIRUV}) tells us that, in IR, the $(-1)^FW$ line should have

    \[ c_{(-1)^FW,(-1)^FW}^\text{IR}c_{(-1)^FW,(-1)^FW}^\text{IR}=e^{-\frac{2\pi i}5}id_1\oplus e^{\frac{4\pi i}5}id_W. \]
    The identity-channel says the corresponding primary should have conformal dimension $h_{(-1)^FW}^\text{IR}$ with
    \[ \theta_{(-1)^FW}^{-2}=e^{-4\pi ih_{(-1)^FW}^\text{IR}}=e^{-\frac{2\pi i}5}, \]
    or
    \begin{equation}
        h_{(-1)^FW}^\text{IR}=\frac1{10}\quad(\text{mod }\frac12).\label{hFWm=5}
    \end{equation}
    Assuming the unitarity, we get a prediction
    \[ h_{(-1)^FW}^\text{IR}=\frac1{10},\frac35,\frac{11}{10},\dots\ . \]
    Indeed, physically expected smallest candidate $h_{(-1)^FW}^\text{IR}=\frac1{10}$ is realized by the primary $|\ep_{\frac1{10}}|^2$ in $m=4$ theory. (Since $h^\text{UV}=\frac25$, the ``monotonicity'' (\ref{hUV>hIR}) forbids $\frac35$.) Similarly, let us look at the double braiding of $W$ with itself. Since the double braidings are the same, we get the same prediction
    \begin{equation}
        h_W^\text{IR}=\frac1{10}\quad(\text{mod }\frac12).\label{hWm=5}
    \end{equation}
    Assuming the unitarity, we get
    \[ h_W^\text{IR}=\frac1{10},\frac35,\frac{11}{10},\dots\ . \]
    Indeed, the next smallest candidate $h_W^\text{IR}=\frac35$ is realized by $|\ep_{\frac35}|^2$ in $m=4$ theory. One can convince oneself that the $m=4$ fermionic minimal model is the only possible IR CFT.
    
    Let us now re-examine this massless flow from the viewpoint of symmetry enhancement. Taking the (quantum) trace of (\ref{DB5}), we obtain (unnormalized) topological $S$-matrix
    \[ \widetilde S_\text{top}=\begin{pmatrix}1&1&\zeta&\zeta\\1&1&\zeta&\zeta\\\zeta&\zeta&-1&-1\\\zeta&\zeta&-1&-1\end{pmatrix}. \]
    The monodromy charge matrix is thus given by
    \begin{equation}
        M=\begin{pmatrix}1&1&1&1\\1&1&1&1\\1&1&-\frac1{\zeta^2}&-\frac1{\zeta^2}\\1&1&-\frac1{\zeta^2}&-\frac1{\zeta^2}\end{pmatrix}.\label{M5}
    \end{equation}
    One finds $(-1)^F$ is transparent. The rank four BFC does not have invertible topological $S$-matrix, and our proposal claims the symmetry should enhance. Indeed, the enhancement is consistent with the RG flow to the $m=4$ theory with eight TDLs:
    \begin{equation}
    \begin{array}{ccccc}
    \text{UV}:&1&(-1)^F&W&(-1)^FW\\
    &\downarrow&\downarrow&\downarrow&\downarrow\\
    \text{IR}:&1&(-1)^F&W&(-1)^FW
    \end{array}.\label{matchingm=5}
    \end{equation}
    Notice that neither of the non-invertible lines, $W$ and $(-1)^FW$, are preserved in the previous example, consistent with the reality condition.
    \item $m=6$:\\
    The theory has 15 TDLs generated by seven primitive lines
    \[ \{(-1)^F,\mcal L_{\frac17,\frac17},\mcal L_{\frac57,\frac57},\mcal L_{\frac43,\frac43},\mcal L_{\frac38,\frac{23}8},\mcal L_{\frac1{56},\frac{85}{56}},\mcal L_{\frac5{56},\frac{33}{56}}\}. \]
    The deformation operator $|\phi_{1,3}|^2=|\ep_{\frac57}|^2$ commutes with five TDLs
    \[ \{1,(-1)^F,\mcal L_{\frac43,\frac43},\mcal L_{\frac38,\frac{23}8},\mcal L_{\frac{23}8,\frac38}=(-1)^F\mcal L_{\frac38,\frac{23}8}\}. \]
    Can this flow be massless? Let us first consider this point in order to justify our assumption of rationality. The $c$-theorem allows fermionic and bosonic minimal models with $m=3,4,5$. Using the absence of simple lines with quantum dimension two, we can first rule out $m=3,4$ CFTs. The spin constraint further rules out bosonic $m=5$ minimal model. Thus we are left with the fermionic $m=5$ minimal model. As we will see below, the surviving five TDLs can be matched with $\{1,(-1)^F,M,N,(-1)^FN\}$. In the previous example, we have seen that none of the three non-invertible lines are preserved. Therefore, the double braiding relation does not rule out the massless scenario. The relation also claims non-invertible lines preserved in the previous example, i.e., $W,(-1)^FW$, should be emergent.
    
    Let us elaborate on the symmetry enhancement. The double braidings of the surviving lines are given by
    \begin{equation}
        \hspace{-70pt}\begin{pmatrix}id_1&id_{(-1)^F}&id_{\mcal L_{\frac43,\frac43}}&id_{\mcal L_{\frac38,\frac{23}8}}&id_{\mcal L_{\frac{23}8,\frac38}}\\id_{(-1)^F}&id_1&id_{\mcal L_{\frac43,\frac43}}&-id_{\mcal L_{\frac{23}8,\frac38}}&-id_{\mcal L_{\frac38,\frac{23}8}}\\id_{\mcal L_{\frac43,\frac43}}&id_{\mcal L_{\frac43,\frac43}}&e^{\frac{2\pi i}3}id_1\oplus e^{\frac{2\pi i}3}id_{(-1)^F}\oplus e^{-\frac{2\pi i}3}id_{\mcal L_{\frac43,\frac43}}&e^{-\frac{2\pi i}3}id_{\mcal L_{\frac38,\frac{23}8}}\oplus e^{\frac{\pi i}3}id_{\mcal L_{\frac{23}8,\frac38}}&e^{\frac{\pi i}3}id_{\mcal L_{\frac38,\frac{23}8}}\oplus e^{-\frac{2\pi i}3}id_{\mcal L_{\frac{23}8,\frac38}}\\id_{\mcal L_{\frac38,\frac{23}8}}&-id_{\mcal L_{\frac{23}8,\frac38}}&e^{-\frac{2\pi i}3}id_{\mcal L_{\frac38,\frac{23}8}}\oplus e^{\frac{\pi i}3}id_{\mcal L_{\frac{23}8,\frac38}}&i\cdot id_1\oplus e^{-\frac{5\pi i}6}id_{\mcal L_{\frac43,\frac43}}&-i\cdot id_{(-1)^F}\oplus e^{\frac{\pi i}6}id_{\mcal L_{\frac43,\frac43}}\\id_{\mcal L_{\frac{23}8,\frac38}}&-id_{\mcal L_{\frac38,\frac{23}8}}&e^{\frac{\pi i}3}id_{\mcal L_{\frac38,\frac{23}8}}\oplus e^{-\frac{2\pi i}3}id_{\mcal L_{\frac{23}8,\frac38}}&-i\cdot id_{(-1)^F}\oplus e^{\frac{\pi i}6}id_{\mcal L_{\frac43,\frac43}}&i\cdot id_1\oplus e^{-\frac{5\pi i}6}id_{\mcal L_{\frac43,\frac43}}\end{pmatrix}.\label{DB6}
    \end{equation}
    Its (quantum) trace gives unnormalized topological $S$-matrix
    \[ \widetilde S_\text{top}=\begin{pmatrix}1&1&2&\sqrt3&\sqrt3\\1&1&2&-\sqrt3&-\sqrt3\\2&2&-\sqrt3&0&0\\\sqrt3&-\sqrt3&0&-\sqrt3&\sqrt3\\\sqrt3&-\sqrt3&0&\sqrt3&-\sqrt3\end{pmatrix}. \]
    Dividing with quantum dimensions, we get monodromy charge matrix
    \begin{equation}
        M=\begin{pmatrix}1&1&1&1&1\\1&1&1&-1&-1\\1&1&-\frac{\sqrt3}4&0&0\\1&-1&0&-\frac1{\sqrt3}&\frac1{\sqrt3}\\1&-1&0&\frac1{\sqrt3}&-\frac1{\sqrt3}\end{pmatrix}.\label{M6}
    \end{equation}
    One sees the symmetric centralizer is trivial. Therefore, we cannot say whether the surviving rank five BFC enhances until we study its central charge. Unfortunately, $\Gamma_\theta$-BFCs are poorly classified, let alone their central charges. If our double braiding relation was unavailable, we had to stop here. However, the relation comes to help us; we have already seen the double braiding relation requires $W$ to be emergent (because the line has non-real double braiding with itself, and if it were non-emergent, it contradicted the reality constraint). Since double braidings have more information than topological $S$-matrix, it is not surprising that we could achieve results otherwise unaccessible with double braidings.
    
    Let us further constrain the massless scenario by predicting IR conformal dimensions. We start from double braiding of $\mcal L_{\frac43,\frac43}$ with itself. The $\mcal L_{\frac43,\frac43}$-channel tells us, in IR, the corresponding primary should have conformal dimension $h^\text{IR}$ with
    \[ \theta^{-1}=e^{-2\pi ih^\text{IR}}=e^{\frac{2\pi i}3}, \]
    or
    \[ h^\text{IR}=\frac23\quad(\text{mod }\frac12). \]
    With unitarity, this means $h^\text{IR}=\frac16,\frac23,\frac76,\dots$. The second smallest candidate is realized $h_M^\text{IR}=\frac23$ by $|\ep_{\frac23}|^2$ in the $m=5$ theory. Similarly, next let us study IR conformal dimension corresponding to $\mcal L_{\frac38,\frac{23}8}$. We look at the double braiding of the line with itself. The identity-channel says, in IR, the corresponding primary should have conformal dimension $h^\text{IR}$ with
    \[ \theta^{-2}=e^{-4\pi ih^\text{IR}}=-i, \]
    or
    \[ h^\text{IR}=\frac18\quad(\text{mod }\frac12). \]
    The other line $\mcal L_{\frac{23}8,\frac38}$ gives the same result. These are indeed satisfied by $h=\frac18,\frac{13}8$, or $\sigma_{\frac18}\bar\sigma_{\frac{13}8},\sigma_{\frac{13}8}\bar\sigma_{\frac18}$ in the $m=5$ theory. The ``monotonicity'' (\ref{hUV>hIR}) uniquely fixes the identifications of surviving lines:
    \begin{equation}
    \begin{array}{cccccc}
    \text{UV}:&1&(-1)^F&\mcal L_{\frac43,\frac43}&\mcal L_{\frac38,\frac{23}8}&\mcal L_{\frac{23}8,\frac38}\\
    &\downarrow&\downarrow&\downarrow&\downarrow&\downarrow\\
    \text{IR}:&1&(-1)^F&M&N&(-1)^FN
    \end{array}.\label{matchingm=6}
    \end{equation}
    \item $m=7$:\\
    We find 24 TDLs generated by five primitive lines
    \[ \{(-1)^F,\mcal L_{\frac5{14},\frac5{14}},\mcal L_{\frac97,\frac97},\mcal L_{\frac34,\frac34},\mcal C\}. \]
    The deformation operator $|\phi_{1,3}|^2=|\ep_{\frac34}|^2$ commutes with six TDLs
    \[ \{1,(-1)^F,\mcal L_{\frac5{14},\frac5{14}},\mcal L_{\frac97,\frac97},\mcal L_{\frac{34}7,\frac{34}7}=(-1)^F\mcal L_{\frac5{14},\frac5{14}},\mcal L_{\frac{39}{14},\frac{39}{14}}=(-1)^F\mcal L_{\frac97,\frac97}\}. \]
    As usual, we first consider whether this flow can be massless. The $c$-theorem allows fermionic and bosonic minimal models with $m=3,4,5,6$. The minimal models with $m=3,4,5$ cannot match the quantum dimensions. Furthremore, the spin constraint rules out bosonic theories. Therefore, the only candidate is the $m=6$ fermionic minimal model. The six surviving lines can be matched with $\{1,(-1)^F,\mcal L_{\frac17,\frac17},\mcal L_{\frac{22}7,\frac{22}7},\mcal L_{\frac57,\frac57},\mcal L_{\frac{12}7,\frac{12}7}\}$. In the previous example, we saw the (conjectural) massless RG flow preserves none of the four non-invertible lines. Thus, our double braiding relation allows massless RG flow. The relation also teaches us that the non-invertible lines $\{\mcal L_{\frac43,\frac43},\mcal L_{\frac38,\frac{23}8},\mcal L_{\frac{23}8,\frac38}\}$ in $m=6$ should be emergent. In this case, we can also see the rank six surviving BFC should be enlarged by studying the monodromy charge matrix. The six surviving TDLs have double braidings
    \begin{equation}
        \hspace{-85pt}\scalebox{0.7}{$\begin{pmatrix}id_1&id_{(-1)^F}&id_{\mcal L_{\frac5{14},\frac5{14}}}&id_{\mcal L_{\frac97,\frac97}}&id_{\mcal L_{\frac{34}7,\frac{34}7}}&id_{\mcal L_{\frac{39}{14},\frac{39}{14}}}\\id_{(-1)^F}&id_1&id_{\mcal L_{\frac{34}7,\frac{34}7}}&id_{\mcal L_{\frac{39}{14},\frac{39}{14}}}&id_{\mcal L_{\frac5{14},\frac5{14}}}&id_{\mcal L_{\frac97,\frac97}}\\id_{\mcal L_{\frac5{14},\frac5{14}}}&id_{\mcal L_{\frac{34}7,\frac{34}7}}&e^{\frac{4\pi i}7}id_1\oplus e^{-\frac{6\pi i}7}id_{\mcal L_{\frac97,\frac97}}&e^{-\frac{4\pi i}7}id_{\mcal L_{\frac5{14},\frac5{14}}}\oplus e^{\frac{2\pi i}7}id_{\mcal L_{\frac{39}{14},\frac{39}{14}}}&e^{\frac{4\pi i}7}id_{(-1)^F}\oplus e^{-\frac{6\pi i}7}id_{\mcal L_{\frac{39}{14},\frac{39}{14}}}&e^{-\frac{4\pi i}7}id_{\mcal L_{\frac{34}7,\frac{34}7}}\oplus e^{\frac{2\pi i}7}id_{\mcal L_{\frac97,\frac97}}\\id_{\mcal L_{\frac97,\frac97}}&id_{\mcal L_{\frac{39}{14},\frac{39}{14}}}&e^{-\frac{4\pi i}7}id_{\mcal L_{\frac5{14},\frac5{14}}}\oplus e^{\frac{2\pi i}7}id_{\mcal L_{\frac{39}{14},\frac{39}{14}}}&e^{\frac{6\pi i}7}id_1\oplus e^{-\frac{4\pi i}7}id_{\mcal L_{\frac97,\frac97}}\oplus e^{\frac{4\pi i}7}id_{\mcal L_{\frac{34}7,\frac{34}7}}&e^{-\frac{4\pi i}7}id_{\mcal L_{\frac{34}7,\frac{34}7}}\oplus e^{\frac{2\pi i}7}id_{\mcal L_{\frac97,\frac97}}&e^{\frac{6\pi i}7}id_{(-1)^F}\oplus e^{-\frac{4\pi i}7}id_{\mcal L_{\frac{39}{14},\frac{39}{14}}}\oplus e^{\frac{4\pi i}7}id_{\mcal L_{\frac5{14},\frac5{14}}}\\id_{\mcal L_{\frac{34}7,\frac{34}7}}&id_{\mcal L_{\frac5{14},\frac5{14}}}&e^{\frac{4\pi i}7}id_{(-1)^F}\oplus e^{-\frac{6\pi i}7}id_{\mcal L_{\frac{39}{14},\frac{39}{14}}}&e^{-\frac{4\pi i}7}id_{\mcal L_{\frac{34}7,\frac{34}7}}\oplus e^{\frac{2\pi i}7}id_{\mcal L_{\frac97,\frac97}}&e^{\frac{4\pi i}7}id_1\oplus e^{-\frac{6\pi i}7}id_{\mcal L_{\frac97,\frac97}}&e^{-\frac{4\pi i}7}id_{\mcal L_{\frac5{14},\frac5{14}}}\oplus e^{\frac{2\pi i}7}id_{\mcal L_{\frac{39}{14},\frac{39}{14}}}\\id_{\mcal L_{\frac{39}{14},\frac{39}{14}}}&id_{\mcal L_{\frac97,\frac97}}&e^{-\frac{4\pi i}7}id_{\mcal L_{\frac{34}7,\frac{34}7}}\oplus e^{\frac{2\pi i}7}id_{\mcal L_{\frac97,\frac97}}&e^{\frac{6\pi i}7}id_{(-1)^F}\oplus e^{-\frac{4\pi i}7}id_{\mcal L_{\frac{39}{14},\frac{39}{14}}}\oplus e^{\frac{4\pi i}7}id_{\mcal L_{\frac5{14},\frac5{14}}}&e^{-\frac{4\pi i}7}id_{\mcal L_{\frac5{14},\frac5{14}}}\oplus e^{\frac{2\pi i}7}id_{\mcal L_{\frac{39}{14},\frac{39}{14}}}&e^{\frac{6\pi i}7}id_1\oplus e^{-\frac{4\pi i}7}id_{\mcal L_{\frac97,\frac97}}\oplus e^{\frac{4\pi i}7}id_{\mcal L_{\frac{34}7,\frac{34}7}}\end{pmatrix}.$}\label{DB7}
    \end{equation}
    The unnormalized topological $S$-matrix is thus given by
    \[ \widetilde S_\text{top}=\begin{pmatrix}1&1&\frac{\cos\frac{3\pi}{14}}{\sin\frac\pi7}&\frac1{2\sin\frac\pi{14}}&\frac{\cos\frac{3\pi}{14}}{\sin\frac\pi7}&\frac1{2\sin\frac\pi{14}}\\1&1&\frac{\cos\frac{3\pi}{14}}{\sin\frac\pi7}&\frac1{2\sin\frac\pi{14}}&\frac{\cos\frac{3\pi}{14}}{\sin\frac\pi7}&\frac1{2\sin\frac\pi{14}}\\\frac{\cos\frac{3\pi}{14}}{\sin\frac\pi7}&\frac{\cos\frac{3\pi}{14}}{\sin\frac\pi7}&-\frac1{2\sin\frac\pi{14}}&1&-\frac1{2\sin\frac\pi{14}}&1\\\frac1{2\sin\frac\pi{14}}&\frac1{2\sin\frac\pi{14}}&1&-\frac1{4\sin\frac{3\pi}{14}\sin\frac\pi{14}}&1&-\frac1{4\sin\frac{3\pi}{14}\sin\frac\pi{14}}\\\frac{\cos\frac{3\pi}{14}}{\sin\frac\pi7}&\frac{\cos\frac{3\pi}{14}}{\sin\frac\pi7}&-\frac1{2\sin\frac\pi{14}}&1&-\frac1{2\sin\frac\pi{14}}&1\\\frac1{2\sin\frac\pi{14}}&\frac1{2\sin\frac\pi{14}}&1&-\frac1{4\sin\frac{3\pi}{14}\sin\frac\pi{14}}&1&-\frac1{4\sin\frac{3\pi}{14}\sin\frac\pi{14}}\end{pmatrix}. \]
    In the basis above, the monodromy charge matrix is given by
    \begin{equation}
        M=\begin{pmatrix}1&1&1&1&1&1\\1&1&1&1&1&1\\1&1&-8\sin\frac\pi{14}\sin^2\frac{3\pi}{14}&8\sin^2\frac\pi{14}\sin\frac{3\pi}{14}&-8\sin\frac\pi{14}\sin^2\frac{3\pi}{14}&8\sin^2\frac\pi{14}\sin\frac{3\pi}{14}\\1&1&8\sin^2\frac\pi{14}\sin\frac{3\pi}{14}&-\frac{\sin\frac\pi{14}}{\sin\frac{3\pi}{14}}&8\sin^2\frac\pi{14}\sin\frac{3\pi}{14}&-\frac{\sin\frac\pi{14}}{\sin\frac{3\pi}{14}}\\1&1&-8\sin\frac\pi{14}\sin^2\frac{3\pi}{14}&8\sin^2\frac\pi{14}\sin\frac{3\pi}{14}&-8\sin\frac\pi{14}\sin^2\frac{3\pi}{14}&8\sin^2\frac\pi{14}\sin\frac{3\pi}{14}\\1&1&8\sin^2\frac\pi{14}\sin\frac{3\pi}{14}&-\frac{\sin\frac\pi{14}}{\sin\frac{3\pi}{14}}&8\sin^2\frac\pi{14}\sin\frac{3\pi}{14}&-\frac{\sin\frac\pi{14}}{\sin\frac{3\pi}{14}}\end{pmatrix}.\label{M7}
    \end{equation}
    One sees $(-1)^F$ is transparent. Thus our proposal requires the rank six surviving BFC enhance in order to make the topological $S$-matrix invertible.
    
    Let us elaborate more on the flow by predicting IR conformal dimensions. We start from double braiding of $\mcal L_{\frac5{14},\frac5{14}}$ with itself. The identity-channel predicts
    \[ \theta^{-2}=e^{-4\pi ih^\text{IR}}=e^{-\frac{4\pi i}7}, \]
    or
    \[ h^\text{IR}=\frac17\quad(\text{mod }\frac12). \]
    The $\mcal L_{\frac{34}7,\frac{34}7}$ gives the same prediction because the two lines have the same double braiding with themselves. The predictions are satisfied by $h=\frac17,\frac{22}7$. The ``monotonicity'' (\ref{hUV>hIR}) uniquely fixes the identifications $\mcal L_{\frac5{14},\frac5{14}}\to\mcal L_{\frac17,\frac17},\mcal L_{\frac{34}7,\frac{34}7}\to\mcal L_{\frac{22}7,\frac{22}7}$. Next, let us study $\mcal L_{\frac97,\frac97}$. The identity-channel of the double braiding with itself predicts
    \[ \theta^{-2}=e^{-4\pi ih^\text{IR}}=e^{-\frac{6\pi i}7}, \]
    or
    \[ h^\text{IR}=\frac3{14}\quad(\text{mod }\frac12). \]
    The $\mcal L_{\frac{39}{14},\frac{39}{14}}$ yields the same prediction because the two lines have the same double braiding with themselves. The predictions are satisfied by $h=\frac57,\frac{12}7$. Again, the ``monotonicity'' (\ref{hUV>hIR}) uniquely fixes the identifications $\mcal L_{\frac97,\frac97}\to\mcal L_{\frac57,\frac57},\mcal L_{\frac{39}{14},\frac{39}{14}}\to\mcal L_{\frac{12}7,\frac{12}7}$. One finds the spin constraints are beautifully satisfied by the unique identifications:
    \begin{equation}
    \begin{array}{ccccccc}
    \text{UV}:&1&(-1)^F&\mcal L_{\frac5{14},\frac5{14}}&\mcal L_{\frac97,\frac97}&\mcal L_{\frac{34}7,\frac{34}7}&\mcal L_{\frac{39}{14},\frac{39}{14}}\\
    &\downarrow&\downarrow&\downarrow&\downarrow&\downarrow&\downarrow\\
    \text{IR}:&1&(-1)^F&\mcal L_{\frac17,\frac17}&\mcal L_{\frac57,\frac57}&\mcal L_{\frac{22}7,\frac{22}7}&\mcal L_{\frac{12}7,\frac{12}7}
    \end{array}.\label{matchingm=7}
    \end{equation}
\end{itemize}

\section{Discussion}\label{discussion}
We proposed when and why symmetries enhance in massless RG flows to fermionic RCFTs. Our proposal is essentially the same as in bosonic cases \cite{K21}. Namely, when the surviving BFC has non-invertible topological $S$-matrix, or inconsistent with the $c$-theorem, in NS-NS spin structure, the surviving BFC should be enlarged to a $\Gamma_\theta$-BFC which simultaneously satisfies the two requirements. We exposed this claim to tests. In half of the examples, we could explain symmetry enhancements requiring the invertibility of the topological $S$-matrix. In the other half of the examples, our proposal alone could not explain the enhancement due to tha lack of classifications of $\Gamma_\theta$-BFCs. Here, our double braiding relation comes to rescue. The relation says that surviving lines have the $opposite$ double braidings in UV and IR even though the $R$-symbols are ``protected'' by the Ocneanu rigidity. Employing the relation, we managed to explain symmetry enhancements in all examples.

The double braiding relation has many other applications. We discussed three of them; 1) prediction on IR conformal dimensions, 2) revealing some structures of the theory space, and 3) giving a necessary condition for a flow to be massless. We also found scaling dimensions ``monotonically'' decrease along RG flows. Together with the predictions coming from our double braiding relation, the ``monotonicity'' imposes strong constraints on IR conformal dimensions. Sometimes, we found the two conditions give unique predictions. Since conformal dimensions are fundamental conformal data characterizing CFTs, our method strongly constrains massless flows. We would report other applications in future.

As a preparation to discuss constraints on RG flows, we first clarified (non)invariants under RG flows. Along the way, we found it useful to view (non)invariants as linking in one higher dimension. When viewed in this perspective, one can immediately generalize our discussions to other dimensions. It would be interesting to study (non)invariants associated to non-invertible symmetries in higher dimensions, an active research field initiated by the groundbreaking work \cite{KNY21}. Since we saw emergent non-invertible TDLs are common due to strong constraints from the double braiding relation, research in this direction may also explain emergent symmetries in higher dimensions.

\section*{Acknowledgement}
We thank Jin Chen for discussions. We also thank Yu Nakayama for suggesting the TDL $\mcal C$ with chiral action in $m=7$ fermionic minimal model and teaching us additional consistency with OPE.

\appendix
\setcounter{section}{0}
\renewcommand{\thesection}{\Alph{section}}
\setcounter{equation}{0}
\renewcommand{\theequation}{\Alph{section}.\arabic{equation}}

\section{Symmetry category of unitary minimal models}\label{symcat}
In this appendix, we study symmetry categories of unitary minimal models.

\subsection{Bosonic theories}
We start from recalling two known facts; invertible symmetries of bosonic unitary minimal models and spins of their lines.

First, Ruelle and Verhoeven classified invertible symmetries of bosonic unitary minimal models \cite{RV98}. The result is the following. All but six models have maximal $\mbb Z_2$ symmetries. The two $D$-type models $(A_4,D_4)$ and  $(D_4,A_6)$ (i.e., critical  and tricritical three-state Potts models) have $S_3$ symmetries, and four $E$-type models $(A,E)$ and $(E,A)$ with $E_7,E_8$ do not have invertible symmetries. The symmetries are known to be anomaly-free \cite{CW20}.

Second, Runkel and Watts realized that for theories with maximal $\mbb Z_2$ symmetries, spins of the generating lines are half-integers when $m=3,4$ mod $4$ \cite{RW20}.

Combining these facts, we can specify symmetry categories of bosonic unitary minimal models. First of all, the categories have invertible $S$-matrices. Thus, they are modular tensor categories (MTCs). A $\mbb Z_2$ object $\eta$ with topological twist $\theta_\eta=-1$ is called a fermion. An MTC $\mcal C$ with a fermion $\eta$ is especially called spin modular category (SMC), $(\mcal C,\eta)$. From the known facts, we immediately learn bosonic minimal models with maximal $\mbb Z_2$ symmetries at $m=3,4$ mod $4$ are described by SMCs $(\mcal C,\eta)$ where $\eta$'s are the $\mbb Z_2$ lines. For SMCs, it is known that $\mbb Z_2$ even sectors $\mcal C_0$ are super-modular.\footnote{This fact can be understood by studying indicators. For an object $j$, we define an indicator $\nu_j$ (or ``charge'') by
\[ (-1)^{\nu_j}:=\frac{\left(S_\text{top}\right)_{j\eta}}{\left(S_\text{top}\right)_{j1}}. \]
Namely, even objects have indicators $\nu=0$ and odd objects have $\nu=1$. (These coincide with the label of $\mbb Z_2$ grading.) The even sector $\mcal C_0$ by definition has $\nu=0$. Now, note that the RHS is nothing but the monodromy charge matrix:
\[ \frac{\left(S_\text{top}\right)_{j\eta}}{\left(S_\text{top}\right)_{j1}}=\frac{\left(S_\text{top}\right)_{j\eta}}{\left(S_\text{top}\right)_{j1}d_\eta}=\frac{\left(S_\text{top}\right)_{j\eta}\left(S_\text{top}\right)_{11}}{\left(S_\text{top}\right)_{j1}\left(S_\text{top}\right)_{\eta1}}\equiv M_{j\eta}. \]
By definition, $\mcal C_0$ is a collection of objects with $M_{j\eta}=1$. Thus in $\mcal C_0$, the fermion $\eta$ is transparent.} A super-modular category $\mcal B$ is defined by a unitary pre-modular category with $\mcal B'=\text{sVec}$.\footnote{More generally, without assumptions of unitarity and sphericity, a BFC $\mcal B$ with $\mcal B'=\text{sVec}$ is called a slightly degenerate modular category \cite{ENO08}.} Namely, a category which fails to be modular due to the presence of unique transparent fermion. When $\mcal C_0$ is super-modular, we can further ask whether it is split super-modular. A super-modular category $\mcal B$ is called split super-modular if $\mcal B$ is given by a Deligne product $\mcal B\cong\text{sVec}\boxtimes\mcal D$ with an MTC $\mcal D$. Here, we remember another fact, Theorem 3.2 of \cite{fMTC}. In particular, we exploit the equivalence
\[ \mcal C_0\text{ is split super-modular}\iff\mcal C\text{ contains a modular subcategory }\mcal E\ni\eta\text{ with }\dim\mcal E=4. \]
Since $\{1,\eta\}$ already has dimension two, for $\mcal C$ to have such $\mcal E$, we need two more invertible lines, or a line with quantum dimension $\sqrt2$. The first possibility is ruled out because we saw the maximal invertible symmetries are $\mbb Z_2$. Thus, for $\mcal C_0$ to be split super-modular, we need a line with quantum dimension $\sqrt2$.

A line $N$ with quantum dimension $\sqrt2$ in a theory with maximal $\mbb Z_2$ symmetry is highly constrained. We find three lines $\{1,\eta,N\}$ form the $\mbb Z_2$ Tambara-Yamagami category $\text{TY}(\mbb Z_2,\chi,\nu_2(N))$. To show this, let us study fusion rules of $N$. Firstly, it has trivial fusion with the identity line $1$:
\[ N1=1N=N. \]
By symmetry of fusion coefficients, this implies $NN=1+\cdots$. Since $N$ has quantum dimension $\sqrt2$, only another invertible line can appear in the RHS. In our theories with maximal $\mbb Z_2$ symmetry, the only possibility is $\eta$. Thus we arrive
\[ NN=1+\eta. \]
The second channel also implies $N\eta=N+\cdots$, but evaluating quantum dimensions of both hand sides, we get $\sqrt2=\sqrt2+\cdots$. Thus no other line can appear in the RHS:
\[ N\eta=\eta N=N. \]
These are nothing but the fusion rules of the $\mbb Z_2$ Tambara-Yamagami category. On the other hand, a CFT with $\text{TY}(\mbb Z_2,\chi,\nu_2(N))$ has self-duality under $\mbb Z_2$-gauging. (See e.g. \cite{TW21}.) Luckily, it is known that only four theories $(A_2,A_3),(A_3,A_4),(A_{10},E_6),(E_6,A_{12})$ are self-dual. One can check all four theories have $3\times3$ submatrices (in the basis $\{1,N,\eta\}$)
\[ \widetilde S=\begin{pmatrix}1&\sqrt2&1\\\sqrt2&0&-\sqrt2\\1&-\sqrt2&1\end{pmatrix}, \]
which is invertible. Thus these four theories have split super-modular even sectors.\footnote{All four models have the same rank three Tambara-Yamagami category $\text{TY}(\mbb Z_2,\chi,\nu_2(N))$ with bicharacter $\chi(\eta,\eta)=-1$ and Frobenius-Schur indicator $\nu_2(N)=+1$. Namely, the rank three fusion subcategories in the four theories have the same $F$-symbols.}

In short, we arrive the following specification of symmetry categories. Bosonic unitary minimal models with maximal $\mbb Z_2$ symmetry at $m=3,4$ mod $4$ are described by SMCs $(\mcal C,\eta)$. Among them, only four theories $(A_2,A_3),(A_3,A_4),(A_{10},E_6),(E_6,A_{12})$ have split-super modular even sectors $\mcal C_0$, i.e., $\mcal C_0\cong\text{sVec}\boxtimes\mcal D$ for an MTC $\mcal D$.\footnote{For the first two models, the MTCs are given by $\mcal D=\text{Vec},\text{Fib}$, respectively. The latter two have rank five and six MTCs with fusion rings
\begin{table}[H]
\begin{center}
\begin{tabular}{c|c|c|c|c|c}
&1&$\mcal L_2$&$\mcal L_3$&$\mcal L_4$&$\mcal L_5$\\\hline
1&1&$\mcal L_2$&$\mcal L_3$&$\mcal L_4$&$\mcal L_5$\\\hline
$\mcal L_2$&&$1+\mcal L_2+\mcal L_3$&$\sum_{i=2}^4\mcal L_i$&$\sum_{i=3}^5\mcal L_i$&$\mcal L_4+\mcal L_5$\\\hline
$\mcal L_3$&&&$1+\sum_{i=2}^5\mcal L_i$&$\sum_{i=2}^5\mcal L_i$&$\mcal L_3+\mcal L_4$\\\hline
$\mcal L_4$&&&&$1+\sum_{i=2}^4\mcal L_i$&$\mcal L_2+\mcal L_3$\\\hline
$\mcal L_5$&&&&&$1+\mcal L_2$
\end{tabular}
\end{center}
\end{table}
\hspace{-10pt}and
\begin{table}[H]
\begin{center}
\begin{tabular}{c|c|c|c|c|c|c}
&1&$\mcal L_2$&$\mcal L_3$&$\mcal L_4$&$\mcal L_5$&$\mcal L_6$\\\hline
1&1&$\mcal L_2$&$\mcal L_3$&$\mcal L_4$&$\mcal L_5$&$\mcal L_6$\\\hline
$\mcal L_2$&&$1+\mcal L_2+\mcal L_3$&$\sum_{i=2}^4\mcal L_i$&$\sum_{i=3}^5\mcal L_i$&$\sum_{i=4}^6\mcal L_i$&$\mcal L_5+\mcal L_6$\\\hline
$\mcal L_3$&&&$1+\sum_{i=2}^5\mcal L_i$&$\sum_{i=2}^6\mcal L_i$&$\sum_{i=3}^6\mcal L_i$&$\mcal L_4+\mcal L_5$\\\hline
$\mcal L_4$&&&&$1+\sum_{i=2}^6\mcal L_i$&$\sum_{i=2}^5\mcal L_i$&$\mcal L_3+\mcal L_4$\\\hline
$\mcal L_5$&&&&&$1+\sum_{i=2}^4\mcal L_i$&$\mcal L_2+\mcal L_3$\\\hline
$\mcal L_6$&&&&&&$1+\mcal L_2$
\end{tabular},
\end{center}
\end{table}
\hspace{-10pt}respectively. We explicitly checked the $5\times5$ and $6\times6$ $S$-submatrices are invertible. The rank five and six MTCs have \cite{GK94} $SU(2)_9/\mbb Z_2$ realization with identifications
\[ 1^\text{here}=1^\text{there},\quad\mcal L_2^\text{here}=\mcal L_2^\text{there},\quad\mcal L_3^\text{here}=\mcal L_4^\text{there},\quad\mcal L_4^\text{here}=\mcal L_3^\text{there},\quad\mcal L_5^\text{here}=\mcal L_1^\text{there}, \]
and $SU(2)_{11}/\mbb Z_2$ realization with identifications
\[ 1^\text{here}=1^\text{there},\quad\mcal L_2^\text{here}=\mcal L_2^\text{there},\quad\mcal L_3^\text{here}=\mcal L_4^\text{there},\quad\mcal L_4^\text{here}=\mcal L_5^\text{there},\quad\mcal L_5^\text{here}=\mcal L_3^\text{there},\quad\mcal L_6^\text{here}=\mcal L_1^\text{there}, \]
respectively.} Even sectors of the other SMCs are non-split super modular. The other models with maximal $\mbb Z_2$ symmetry are described by non-spin MTCs.

This leaves the two minimal models with $S_3$ symmetry, i.e., three-state Potts models $(A_4,D_4),(D_4,A_6)$. They are described by MTCs with rank six and nine, respectively. In particular, they are $\mbb Z_3$ paraspin modular categories. (See the Appendix \ref{PMC}.) The MTCs have product structures $\mcal C\cong\text{Vec}_{\mbb Z_3}^1\boxtimes\mcal D$ with another MTC $\mcal D$. They are given by $\text{Fib}$ and a rank three MTC with fusion ring
\begin{table}[H]
\begin{center}
\begin{tabular}{c|c|c|c}
&1&$\mcal L_2$&$\mcal L_3$\\\hline
1&1&$\mcal L_2$&$\mcal L_3$\\\hline
$\mcal L_2$&&$1+\mcal L_3$&$\mcal L_2+\mcal L_3$\\\hline
$\mcal L_3$&&&$1+\mcal L_2+\mcal L_3$
\end{tabular},
\end{center}
\end{table}
\hspace{-10pt}respectively. The latter has a $SU(2)_5/\mbb Z_2$ realization \cite{GK94} with identifications
\[ 1^\text{here}=1^\text{there},\quad\mcal L_2^\text{here}=\mcal L_1^\text{there},\quad\mcal L_3^\text{here}=\mcal L_2^\text{there}. \]

Before we move to fermionic theories, let us briefly mention RG flows from some exceptional models. As we explained in the main part, relevant deformation and discrete gauging commute. Two three-state Potts models are obtained by gauging $\mbb Z_2$ symmetries. More concretely, we have
\[ (A_4,D_4)=M(6,5)/\mbb Z_2,\quad(D_4,A_6)=M(7,6)/\mbb Z_2. \]
Here, recall we have the RG flow $M(7,6)\to M(6,5)$ triggered by $\phi_{1,3}$ in $M(7,6)$. Using the commutativity, we get the RG flow\footnote{Flipping some signs, we would also get RG flows among $\mbb Z_3$ parafermionic theories.}
\[ (D_4,A_6)\to(A_4,D_4). \]
In fact, one can check the primary $\phi_{1,3}$ in $(D_4,A_6)$ commutes with $\mbb Z_3$ line $\eta$. Since the lowest CFT with $\mbb Z_3$ symmetry is $(A_4,D_4)$, the only possible massless flow is the one above. The surviving lines have double braidings
\[ \begin{pmatrix}id_1&id_\eta&id_{\eta^2}\\id_\eta&\omega^2id_{\eta^2}&\omega id_1\\id_{\eta^2}&\omega id_1&\omega^2id_\eta\end{pmatrix}. \]
On the other hand, the rank three $\mbb Z_3$ subcategory of $(A_4,D_4)$ has double braidings
\[ \begin{pmatrix}id_1&id_\eta&id_{\eta^2}\\id_\eta&\omega id_{\eta^2}&\omega^2id_1\\id_{\eta^2}&\omega^2id_1&\omega id_\eta\end{pmatrix}. \]
One sees our double braiding relation is beautifully satisfied. Furthermore, their topological $S$-matrices obey
\[ \left(\widetilde S_\text{top}^{(A_4,D_4)}\right)\Big|_{\mbb Z_3}=\left(\widetilde S_\text{top}^{(D_4,A_6)}\right)\Big|_{\mbb Z_3}^* \]
for the rank three $\mbb Z_3$ subcategory. This example supports our proposal (\ref{topSIRUV}). This flow also supports our ``monotonicity'' (\ref{hUV>hIR}) beyond $\mbb Z_2$ because $\frac43=h^\text{UV}_\eta>h^\text{IR}_\eta=\frac23$.

The flow should have emergent symmetry because the surviving MTC has rank three while the IR theory has rank six. Therefore, we have to explain why the emergent lines appear to make it natural. Fortunately, it is not difficult to give an explanation thanks to the classifications of MTCs. According to \cite{GK94}, the surviving rank three MTC has central charge $2$ mod $4$. With unitarity, this means $c=2,6,10,\dots$, which cannot be smaller than $c_\text{UV}=\frac67$. Thus, emergent lines should appear to make the IR MTC consistent with the $c$-theorem. Can one emergent line make it satisfy both modularity and the $c$-theorem? No. There is no rank four MTC with $\mbb Z_3$ fusion ring \cite{GK94} (see also \cite{rank4,rank5}). How about two emergent lines? Again, the answer is no. According to the same papers, there is no rank five MTC with $\mbb Z_3$ fusion ring, either. How about three emergent lines? This time, the answer is yes. There is only one\footnote{An MTC with $SU(6)_1$ realization also contains $\mbb Z_3$ fusion ring, but it has central charge $c=1,3$ mod $4$, which cannot be smaller than $c_\text{UV}=\frac67$ with unitarity.} suitable rank six MTC with $SU(3)_2$ realization. The identifications are given by
\[ \hspace{-30pt}1^\text{here}=1^\text{there},\quad W^\text{here}=\mcal L_1^\text{there},\quad\eta^\text{here}=\mcal L_2^\text{there},\quad\left(\eta^2\right)^\text{here}=\mcal L_3^\text{there},\quad\left(\eta W\right)^\text{here}=\mcal L_4^\text{there},\quad\left(\eta^2W\right)^\text{here}=\mcal L_5^\text{there}, \]
or
\[ \hspace{-30pt}1^\text{here}=1^\text{there},\quad W^\text{here}=\mcal L_1^\text{there},\quad\eta^\text{here}=\mcal L_3^\text{there},\quad\left(\eta^2\right)^\text{here}=\mcal L_2^\text{there},\quad\left(\eta W\right)^\text{here}=\mcal L_5^\text{there},\quad\left(\eta^2W\right)^\text{here}=\mcal L_4^\text{there}. \]
The two choices are exchanged by charge conjugation. The rank six MTC has central charge $c=\frac{4n+4}5$ with $n<5$, which can be smaller than $c_\text{UV}=\frac67$. From the $c$-theorem, we need $n\le0$, but the assumption of masslss phase requires $0\le n$. We are left with the only possibility $n=0$, and indeed this gives the central charge of the three-state Potts model $c=\frac45$, consistent with the $c$-theorem. (Notice that fusion ring of emergent lines has also been fixed by our two requirements.)

Let us also comment on another flow. A reader may notice that emergent lines we explained are all non-invertible. Can we also explain emergent invertible TDLs in our language? Yes, we can. Pick $M(8,7)/\mbb Z_2$ theory as a UV theory. From the commutativity, we know a relevant operator $\phi_{1,3}$ of the theory triggers massless flow to $M(7,6)/\mbb Z_2=(D_4,A_6)$. The $\mbb Z_3$ line $\eta$ of the IR model should be emergent from the reality condition.

Let us also study relevant deformation of $(E_6,A_{12})$. Since the theory is self-dual under $\mbb Z_2$ gauging, we cannot resort to the commutativity. One finds the $\phi_{1,3}$-deformation only preserves the Tambara-Yamagami category. Assuming massless scenario, we have three lower CFTs containing the rank three MTC, $(A_2,A_3),(A_3,A_4),(A_{10},E_6)$. Since the theories have the same $F$-symbols, we cannot rule out any if we were unaware of the double braiding relation. However, with the relation, we can rule out the critical Ising model $(A_2,A_3)$ because it has the same double braidings as the rank three TY category of $(E_6,A_{12})$. We are not sure which theory, $(A_3,A_4)$ or $(A_{10},E_6)$, can be the IR theory, but the ``naturality'' \cite{K21} (three emergent lines vs. 12 emergent lines) and the principle of the lowest energy ($\frac7{16}$ vs. $\frac{31}{16}$) favor the tricritical Ising model $(A_3,A_4)$.

\subsection{Fermionic theories}
Since symmetry category, or even invertible symmetries of fermionic minimal models\footnote{As we reviewed in the previous subsection, there are two minimal models with anomaly-free $\mbb Z_3$ symmetry, three-state Potts models. One can parafermionize the theories using the symmetry. This was performed in \cite{YF20}. It would be interesting to study their symmetry categories.} are little explored, this subsection will be quite preliminary. Furthermore, we limit ourselves to NS-NS spin structure for simplicity.

For fermionic minimal models obtained via fermionization of $M(m+1,m)$ at $m=3,4$ mod $4$, we can prove the following; the theories have $\mbb Z_2\times\mbb Z_2$ fusion ring. This can be shown by studying fusion rules of primaries in the $V_1^1=V$ sector.

Let us start from $m=3$ mod $4$, or $m=4M+3$ with $M=0,1,\dots$ . Following the notation of \cite{HNT}, we have $q=m+1=4M+4$ and $p=m=4M+3$. Accordingly, the $V_1^1=V$ sector has $r\equiv\frac q2+1\equiv1$, or $r=1,3,\dots,q-1=4M+3$. In particular, the sector has primaries with $(r,s)=(1,1),(4M+3,1)$. These Kac indices give $\phi_{1,1}\bar\phi_{4M+3,1}=\bar\ep_{\frac12+M(4M+3)},\phi_{4M+3,1}\bar\phi_{1,1}=\ep_{\frac12+M(4M+3)}$, respectively.\footnote{Primaries with odd $s$ are even under $\eta$.} Now, let us study fusion rules of these primaries with itself. One way to compute fusion rules in fermionic minimal models is as follows. Calculate fusions in holomorphic and anti-holomorphic sectors separetely, and keep terms (consistent with the fermion parity) only if the combination of holomorphic and anti-holomorphic sector exists in the theory. Since we know
\[ \ep_{\frac12+M(4M+3)}\times\ep_{\frac12+M(4M+3)}=id,\quad\bar\ep_{\frac12+M(4M+3)}\times\bar\ep_{\frac12+M(4M+3)}=\bar{id}, \]
the two primaries obey $\mbb Z_2\times\mbb Z_2$ fusion rules.

The case $m=4$ mod $4$, or $m=4M+4$ with $M=0,1,\dots$ proceed in exactly the same way, so we do not repeat the details. The key primaries are given by $\phi_{1,1}\bar\phi_{4M+3,1}=\bar\ep_{\frac32+M(4M+5)},\phi_{4M+3,1}\bar\phi_{1,1}=\ep_{\frac32+M(4M+5)}$ in this case.\footnote{Primaries with odd $r$ are even under $\eta$.}

One may expect the two primaries give $\mbb Z_2\times\mbb Z_2$ TDLs. Indeed, in $m=3,4$, we found two $\mbb Z_2$ symmetries generated by $(-1)^F$ and another q-type $\mbb Z_2$ line. However, in $m=7$ (and $m=8$), we do not have additional $\mbb Z_2$ symmetry. It seems Kramers-Wannier duality line $N$ in bosonic minimal model ``splits'' to two q-type $\mbb Z_2$ lines. (We also see this ``splitting'' in two self-dual exceptional minimal models \cite{K20}.)

One may wonder what happens in the other cases, $m=5,6$ mod $4$. In these cases, we conjecture all TDLs are m-type, and the (topological) $S$-matrices are the same (up to possible basis change) in bosonic and fermionic minimal models (in NS-NS spin structure). In fact, we find the two properties hold when $m=5,6$.

As a necessary condition, it is not hard to show the following: bosonic and fermionic minimal models (in NS-NS spin structure) have exactly the same Kac indices. This can be easily seen by studying the decomposition into four sectors. Since $V_0^0=S$ sectors are common, we only have to compare $V_0^1=T$ and $V_1^1=V$ sectors. When $m=5$ mod $4$, or $m=4M+1$ with $M=1,2,\dots$, $q=m+1=4M+2$ and $p=m=4M+1$. Thus, the defining condition of the $V_1^1=V$ sector reduces to $r\equiv\frac q2+1\equiv0$, the same as the $V_0^1=T$ sector. In case of $m=6$ mod $4$, or $m=4M+2$ with $M=1,2,\dots,$ $p=m+1=4M+3$ and $q=m=4M+2$. Hence, the defining condition of the $V_1^1=V$ sector reduces to $r\equiv\frac q2+1\equiv0$, the same as that of the $V_0^1=T$ sector. One also finds the only difference between the two sectors is that the anti-holomorphic sectors are ``twisted'' by the fusion with the primary corresponding to the $\mbb Z_2$ line. We will abuse the notation, and denote the twisted $\phi_i$ primary as $\phi_{\eta i}$. Then we can show primaries in bosonic and fermionic minimal models (in NS-NS spin structure) obey the same fusion rule. To show this, we perform case analysis.

First, let us consider the case two operators belong to the same sector. Then, fermion parity conservation requires the resulting operators to be bosonic. In case both initial operators belong to the $V_0^0=S$ sector, we thus have
\[ \phi_i\bar\phi_i\times\phi_j\bar\phi_j=\sum_{k,l}\left(N_{ij}^k\phi_k\right)\left(N_{ij}^l\bar\phi_l\right)\Big|=\sum_kN_{ij}^kN_{ij}^k\phi_k\bar\phi_k, \]
where $\Big|$ means the truncation explained above. Since the fusion coefficients are $0$ or $1$ in the diagonal bosonic minimal models, this is the same as in the bosonic minimal model. In case both initial operators belong to the $V_1^1=V$ sector, we have
\[ \phi_i\bar\phi_{\eta i}\times\phi_j\bar\phi_{\eta j}=\sum_{k,l}\left(N_{ij}^k\phi_k\right)\left(N_{ij}^l\bar\phi_l\right)\Big|=\sum_kN_{ij}^kN_{ij}^k\phi_k\bar\phi_k. \]
Again, this is the same as in the bosonic minimal model.

Finally, let us consider the case two operators belong to different sectors. Then, fermion parity conservation requires the resulting operators to be fermionic. Thus we have
\[ \phi_i\bar\phi_i\times\phi_j\bar\phi_{\eta j}=\sum_{k,l}\left(N_{ij}^k\phi_k\right)\left(N_{ij}^l\bar\phi_{\eta l}\right)\Big|=\sum_kN_{ij}^kN_{ij}^k\phi_k\bar\phi_{\eta k}. \]
Therefore, we find bosonic and fermionic minimal models (in NS-NS spin structure) obey the same fusion rules.

Fusion rules are computed from $S$-matrices via the Verlinde formula. In addition, our experience tells that actions of TDLs are also the same. This, in the first place, means quantum dimensions are the same. The matching of quantum dimensions furthermore would imply $S$-matrices are the same. These evidences strongly support $S$-matrices are the same in bosonic and fermionic minimal models at $m=5,6$ mod $4$.

\section{TDLs in fermionic minimal models}\label{TDLs}
In this appendix, we list necessary data of TDLs to understand our discussion in the body of this paper. To find TDLs, we employ the (modified) Cardy condition \cite{C86,KCXC}. Since the method was reviewed/explained in the appendix A of \cite{KCXC}, we will be brief here.

To find a TDL $\mcal L$, one postulates an action on primaries encoded in a mass matrix $M_{\mcal L}$. Then one performs modular $S$-transformations. This gives a trace over a defect Hilbert space $\mcal H_{\mcal L}$. For the space to have a physical interpretation, the expansion in terms of characters should take values in a suitable space.

In fermionic theories, there are two types of lines, m-type and q-type \cite{ALW17}. The former obeys the ordinary cardy condition
\begin{equation}
    (SM_{\mcal L}S^\dagger)_{ij}\stackrel!\in\mbb N,\label{Cardyc}
\end{equation}
while the latter obeys the modified Cardy condition
\begin{equation}
    (SM_{\mcal L}S^\dagger)_{ij}\stackrel!\in\sqrt2\mbb N.\label{modCardyc}
\end{equation}
Note that the matrix $S$ is the modular $S$-matrix, and not the topological $S$-matrix $S_\text{top}$.

For each $m$, we solve these conditions with a help of computer. Since the maximum number of TDLs are given by that of primaries, we can stop when we find that number of TDLs. In all cases, we find the number of TDLs is the same as the number of primaries. Sometimes, we find more solutions to the (modified) Cardy conditions. To find genuine TDLs, we should also require the following consistency condition. Given two TDLs $\mcal L_1,\mcal L_2$, we can act them on an operator $\phi$ in two ways, either act them one after another $\mcal L_1(\mcal L_2|\phi\ra)$, or first fuse the two TDLs and act the resulting TDL $(\mcal L_1\mcal L_2)|\phi\ra$. The two methods should be consistent. In particular, if $\mcal L_1$ and $\mcal L_2$ are genuine TDLs, then the fused $(\mcal L_1\mcal L_2)$ should also solve the (modified) Cardy condition. In $m=7$ fermionic minimal model, this consistency condition rules out $\mcal L$ we found in v1.

One could solve the (modified) Cardy conditions by a brute-force search, but experience tells us holomorphic and anti-holomorphic sectors combine only if they have the same actions in bosonic theories. This ansatz makes the computation much faster. Below, we list TDLs found in this way. We also present relevant spin contents associated to their defect Hilbert spaces.

\subsection{$m=3$}
The theory has four TDLs (two m-type and two q-type) generated by two primitive lines $\{(-1)^F,(-1)^{F_R}\}$. They act as
\begin{table}[H]
\begin{center}
\begin{tabular}{c|c|c||c|c}
	NS-NS&$id$&$\ep$&$\psi$&$\bar\psi$\\\hline
	$\widehat{(-1)^F}$&$1$&$1$&$-1$&$-1$\\
    $\widehat{(-1)^{F_R}}$&$1$&$-1$&1&$-1$
\end{tabular}.
\end{center}
\end{table}
Their associated spin contents are given by
\begin{align*}
    \mcal H_{(-1)^F}:&\quad s\in\{0\}\text{ mod }1,\\
    \mcal H_{(-1)^{F_R}}:&\quad s\in\{-\frac1{16},\frac7{16}\}\text{ mod }1.
\end{align*}

\subsection{$m=4$}
The theory has eight TDLs (four m-type and four q-type) generated by three primitive lines $\{(-1)^F,W,R\}$. They act as
\begin{table}[H]
\begin{center}
\begin{tabular}{c|c|c|c|c||c|c|c|c}
	NS-NS&$id$&$|\ep|^2$&$|\ep'|^2=|\phi_{1,3}|^2$&$|\ep''|^2$&$\ep''$&$\bar\ep''$&$\ep\bar\ep'$&$\ep'\bar\ep$\\\hline
	$\widehat{(-1)^F}$&$1$&$1$&$1$&$1$&$-1$&$-1$&$-1$&$-1$\\
	$\hat W$&$\zeta$&$-\zeta^{-1}$&$-\zeta^{-1}$&$\zeta$&$\zeta$&$\zeta$&$-\zeta^{-1}$&$-\zeta^{-1}$\\
	$\hat R$&$1$&$-1$&$1$&$-1$&$-1$&$1$&$-1$&$1$
\end{tabular},
\end{center}
\end{table}
\hspace{-10pt}where $\zeta:=\frac{1+\sqrt5}2$ is the golden ratio. The relevant spin contents are given by
\begin{align*}
    \mcal H_{(-1)^F}:&\quad s\in\{0\}\text{ mod }1,\\
    \mcal H_W:&\quad s\in\{0,\pm\frac25,\pm\frac1{10}\}\text{ mod }1,\\
    \mcal H_{(-1)^FW}:&\quad s\in\{0,\pm\frac25\}\text{ mod }1,\\
    \mcal H_R:&\quad s\in\{-\frac1{16},\frac7{16}\}\text{ mod }1.
\end{align*}

\subsection{$m=5$}
The theory has 10 TDLs (all of them are m-type) generated by four primitive lines $\{(-1)^F,M,W,N\}$. They act as
\begin{table}[H]
\begin{center}
\begin{tabular}{c|c|c|c|c|c|c|c|c|c|c}
	NS-NS&$id$&$\sigma'\bar\sigma'''$&$|\ep''|^2=|\phi_{1,3}|^2$&$\sigma'''\bar\sigma'$&$|\ep''''|^2$&$|\ep'|^2$&$\sigma\bar\sigma''$&$|\ep|^2$&$\sigma''\bar\sigma$&$|\ep'''|^2$\\\hline
	$\widehat{(-1)^F}$&$1$&$-1$&$1$&$-1$&$1$&$1$&$-1$&$1$&$-1$&$1$\\
	$\hat M$&$2$&$0$&$-1$&$0$&$2$&$2$&$0$&$-1$&$0$&$2$\\
	$\hat W$&$\zeta$&$\zeta$&$\zeta$&$\zeta$&$\zeta$&$-\zeta^{-1}$&$-\zeta^{-1}$&$-\zeta^{-1}$&$-\zeta^{-1}$&$-\zeta^{-1}$\\
	$\hat N$&$\sqrt3$&$1$&$0$&$-1$&$-\sqrt3$&$-\sqrt3$&$-1$&$0$&$1$&$\sqrt3$
\end{tabular}.
\end{center}
\end{table}
The relevant spin contents are given by
\begin{align*}
    \mcal H_{(-1)^F}:&\quad s\in\{0\}\text{ mod }1,\\
    \mcal H_M:&\quad s\in\{0,\pm\frac12,\pm\frac13\}\text{ mod }1,\\
    \mcal H_W:&\quad s\in\{0,\pm\frac12,\pm\frac25,\pm\frac1{10}\}\text{ mod }1,\\
    \mcal H_{(-1)^FW}:&\quad s\in\{0,\pm\frac25\}\text{ mod }1,\\
    \mcal H_N,\mcal H_{(-1)^FN}:&\quad s\in\{\pm\frac1{24},\pm\frac18,\pm\frac38,\pm\frac{11}{24}\}\text{ mod }1.
\end{align*}

\subsection{$m=6$}
The theory has 15 TDLs (all of them are m-type) generated by seven primitive lines
\[ \{(-1)^F,\mcal L_{\frac17,\frac17},\mcal L_{\frac57,\frac57},\mcal L_{\frac43,\frac43},\mcal L_{\frac38,\frac{23}8},\mcal L_{\frac1{56},\frac{85}{56}},\mcal L_{\frac5{56},\frac{33}{56}}\}. \]
They act as
\begin{table}[H]
\hspace{-50pt}
\scalebox{0.6}{\begin{tabular}{c|c|c|c|c|c|c|c|c|c||c|c|c|c|c|c}
	NS-NS&$id$&$|\ep_{\frac17}|^2$&$|\ep_{\frac57}|^2=|\phi_{1,3}|^2$&$|\ep_{\frac43}|^2$&$|\ep_{\frac{10}{21}}|^2$&$|\ep_{\frac1{21}}|^2$&$|\ep_5|^2$&$|\ep_{\frac{22}7}|^2$&$|\ep_{\frac{12}7}|^2$&$\sigma_{\frac38}\bar\sigma_{\frac{23}8}$&$\sigma_{\frac{23}8}\bar\sigma_{\frac38}$&$\sigma_{\frac1{56}}\bar\sigma_{\frac{85}{56}}$&$\sigma_{\frac{85}{56}}\bar\sigma_{\frac1{56}}$&$\sigma_{\frac5{56}}\bar\sigma_{\frac{33}{56}}$&$\sigma_{\frac{33}{56}}\bar\sigma_{\frac5{56}}$\\\hline
	$\widehat{(-1)^F}$&$1$&$1$&$1$&$1$&$1$&$1$&$1$&$1$&$1$&$-1$&$-1$&$-1$&$-1$&$-1$&$-1$\\
	$\hat{\mcal L}_{\frac17,\frac17}$&$\frac{\cos\frac{3\pi}{14}}{\sin\frac\pi7}$&$-2\sin\frac{3\pi}{14}$&$2\sin\frac\pi{14}$&$\frac{\cos\frac{3\pi}{14}}{\sin\frac\pi7}$&$-2\sin\frac{3\pi}{14}$&$2\sin\frac\pi{14}$&$\frac{\cos\frac{3\pi}{14}}{\sin\frac\pi7}$&$-2\sin\frac{3\pi}{14}$&$2\sin\frac\pi{14}$&$2\cos\frac\pi7$&$2\cos\frac\pi7$&$-2\sin\frac{3\pi}{14}$&$-2\sin\frac{3\pi}{14}$&$2\sin\frac\pi{14}$&$2\sin\frac\pi{14}$\\
	$\hat{\mcal L}_{\frac57,\frac57}$&$\frac1{2\sin\frac\pi{14}}$&$\frac{\sin\frac\pi7}{\cos\frac{3\pi}{14}}$&$-\frac1{2\sin\frac{3\pi}{14}}$&$\frac1{2\sin\frac\pi{14}}$&$\frac{\sin\frac\pi7}{\cos\frac{3\pi}{14}}$&$-\frac1{2\sin\frac{3\pi}{14}}$&$\frac1{2\sin\frac\pi{14}}$&$\frac{\sin\frac\pi7}{\cos\frac{3\pi}{14}}$&$-\frac1{2\sin\frac{3\pi}{14}}$&$-\frac1{2\sin\frac\pi{14}}$&$-\frac1{2\sin\frac\pi{14}}$&$-\frac{\sin\frac\pi7}{\cos\frac{3\pi}{14}}$&$-\frac{\sin\frac\pi7}{\cos\frac{3\pi}{14}}$&$\frac1{2\sin\frac{3\pi}{14}}$&$\frac1{2\sin\frac{3\pi}{14}}$\\
	$\hat{\mcal L}_{\frac43,\frac43}$&$2$&$2$&$2$&$-1$&$-1$&$-1$&$2$&$2$&$2$&$0$&$0$&$0$&$0$&$0$&$0$\\
	$\hat{\mcal L}_{\frac38,\frac{23}8}$&$\sqrt3$&$-\sqrt3$&$\sqrt3$&$0$&$0$&$0$&$-\sqrt3$&$\sqrt3$&$-\sqrt3$&$-1$&$1$&$1$&$-1$&$-1$&$1$\\
	$\hat{\mcal L}_{\frac1{56},\frac{85}{56}}$&$\frac{\sqrt3\cos\frac{3\pi}{14}}{\sin\frac\pi7}$&$2\sqrt3\sin\frac{3\pi}{14}$&$2\sqrt3\sin\frac\pi{14}$&$0$&$0$&$0$&$-2\sqrt3\cos\frac\pi7$&$-2\sqrt3\sin\frac{3\pi}{14}$&$-2\sqrt3\sin\frac\pi{14}$&$\frac{\cos\frac{3\pi}{14}}{\sin\frac\pi7}$&$-2\cos\frac\pi7$&$2\sin\frac{3\pi}{14}$&$-2\sin\frac{3\pi}{14}$&$2\sin\frac\pi{14}$&$-2\sin\frac\pi{14}$\\
	$\hat{\mcal L}_{\frac5{56},\frac{33}{56}}$&$\frac{\sqrt3}{2\sin\frac\pi{14}}$&$-\frac{\sqrt3\sin\frac\pi7}{\cos\frac{3\pi}{14}}$&$-\frac{\sqrt3}{2\sin\frac{3\pi}{14}}$&$0$&$0$&$0$&$-\frac{\sqrt3}{2\sin\frac\pi{14}}$&$\frac{\sqrt3\sin\frac\pi7}{\cos\frac{3\pi}{14}}$&$\frac{\sqrt3}{2\sin\frac{3\pi}{14}}$&$-\frac1{2\sin\frac\pi{14}}$&$\frac1{2\sin\frac\pi{14}}$&$\frac{\sin\frac\pi7}{\cos\frac{3\pi}{14}}$&$-\frac{\sin\frac\pi7}{\cos\frac{3\pi}{14}}$&$\frac1{2\sin\frac{3\pi}{14}}$&$-\frac1{2\sin\frac{3\pi}{14}}$
\end{tabular}.}
\end{table}
The relevant spin contents are given by
\begin{align*}
    \mcal H_{(-1)^F}:&\quad s\in\{0\}\text{ mod }1,\\
    \mcal H_{\mcal L_{\frac17,\frac17}}:&\quad s\in\{0,\pm\frac1{14},\pm\frac17,\pm\frac5{14},\pm\frac37,\pm\frac12\}\text{ mod }1,\\
    \mcal H_{\mcal L_{\frac57,\frac57}}:&\quad s\in\{0,\pm\frac27,\pm\frac37\}\text{ mod }1,\\
    \mcal H_{\mcal L_{\frac43,\frac43}}:&\quad s\in\{0,\pm\frac13,\pm\frac12\}\text{ mod }1,\\
    \mcal H_{\mcal L_{\frac38,\frac{23}8}},\mcal H_{\mcal L_{\frac{23}8,\frac38}}:&\quad s\in\{\pm\frac1{24},\pm\frac18,\pm\frac38,\pm\frac{11}{24}\}\text{ mod }1,\\
    \mcal H_{\mcal L_{\frac{22}7,\frac{22}7}}:&\quad s\in\{0,\pm\frac17,\pm\frac37\}\text{ mod }1,\\
    \mcal H_{\mcal L_{\frac{12}7,\frac{12}7}}:&\quad s\in\{0,\pm\frac1{14},\pm\frac3{14},\pm\frac27,\pm\frac37,\pm\frac12\}\text{ mod }1.
\end{align*}

\subsection{$m=7$}
The theory has 24 TDLs (all m-type) generated by five primitive lines\footnote{The presence of the non-invertible TDL $\mcal C$ with quantum dimension $\sqrt{2+\sqrt2}$ was suggested by Yu Nakayama. He also taught us additional consistency with OPE, which rules out q-type $\mbb Z_2$ TDL $\mcal L$ we found in v1. We thank Yu Nakayama for his observations.}
\[ \{(-1)^F,\mcal L_{\frac5{14},\frac5{14}},\mcal L_{\frac97,\frac97},\mcal L_{\frac34,\frac34},\mcal C\}. \]
They act as
\begin{landscape}
\begin{table}[H]
\hspace{-50pt}
\scalebox{0.5}{\begin{tabular}{c||c|c|c|c|c|c|c|c|c|c|c|c||c|c|c|c|c|c|c|c|c|c|c|c}
	NS-NS&$id$&$|\ep_{\frac5{14}}|^2$&$|\ep_{\frac97}|^2$&$|\ep_{\frac34}|^2=|\phi_{1,3}|^2$&$|\ep_{\frac3{28}}|^2$&$|\ep_{\frac1{28}}|^2$&$|\ep_{\frac{13}4}|^2$&$|\ep_{\frac{45}{28}}|^2$&$|\ep_{\frac{15}{28}}|^2$&$|\ep_{\frac{15}2}|^2$&$|\ep_{\frac{34}7}|^2$&$|\ep_{\frac{39}{14}}|^2$&$\bar\ep_{\frac{15}2}$&$\ep_{\frac{15}2}$&$\ep_{\frac5{14}}\bar\ep_{\frac{34}7}$&$\ep_{\frac{34}7}\bar\ep_{\frac5{14}}$&$\ep_{\frac97}\bar\ep_{\frac{39}{14}}$&$\ep_{\frac{14}{39}}\bar\ep_{\frac97}$&$\ep_{\frac34}\bar\ep_{\frac{13}4}$&$\ep_{\frac{13}4}\bar\ep_{\frac34}$&$\ep_{\frac3{28}}\bar\ep_{\frac{45}{28}}$&$\ep_{\frac{45}{28}}\bar\ep_{\frac3{28}}$&$\ep_{\frac1{28}}\bar\ep_{\frac{15}{28}}$&$\ep_{\frac{15}{28}}\bar\ep_{\frac1{28}}$\\\hline
	$\widehat{(-1)^F}$&$1$&$1$&$1$&$1$&$1$&$1$&$1$&$1$&$1$&$1$&$1$&$1$&$-1$&$-1$&$-1$&$-1$&$-1$&$-1$&$-1$&$-1$&$-1$&$-1$&$-1$&$-1$\\
	$\hat{\mcal L}_{\frac5{14},\frac5{14}}$&$\frac{\cos\frac{3\pi}{14}}{\sin\frac\pi7}$&$-2\sin\frac{3\pi}{14}$&$2\sin\frac\pi{14}$&$\frac{\cos\frac{3\pi}{14}}{\sin\frac\pi7}$&$-2\sin\frac{3\pi}{14}$&$2\sin\frac\pi{14}$&$\frac{\cos\frac{3\pi}{14}}{\sin\frac\pi7}$&$-2\sin\frac{3\pi}{14}$&$2\sin\frac\pi{14}$&$\frac{\cos\frac{3\pi}{14}}{\sin\frac\pi7}$&$-2\sin\frac{3\pi}{14}$&$2\sin\frac\pi{14}$&$\frac{\cos\frac{3\pi}{14}}{\sin\frac\pi7}$&$\frac{\cos\frac{3\pi}{14}}{\sin\frac\pi7}$&$-2\sin\frac{3\pi}{14}$&$-2\sin\frac{3\pi}{14}$&$2\sin\frac\pi{14}$&$2\sin\frac\pi{14}$&$\frac{\cos\frac{3\pi}{14}}{\sin\frac\pi7}$&$\frac{\cos\frac{3\pi}{14}}{\sin\frac\pi7}$&$-2\sin\frac{3\pi}{14}$&$-2\sin\frac{3\pi}{14}$&$2\sin\frac\pi{14}$&$2\sin\frac\pi{14}$\\
	$\hat{\mcal L}_{\frac97,\frac97}$&$\frac1{2\sin\frac\pi{14}}$&$\frac{\sin\frac\pi7}{\cos\frac{3\pi}{14}}$&$-\frac1{2\sin\frac{3\pi}{14}}$&$\frac1{2\sin\frac\pi{14}}$&$\frac{\sin\frac\pi7}{\cos\frac{3\pi}{14}}$&$-\frac1{2\sin\frac{3\pi}{14}}$&$\frac1{2\sin\frac\pi{14}}$&$\frac{\sin\frac\pi7}{\cos\frac{3\pi}{14}}$&$-\frac1{2\sin\frac{3\pi}{14}}$&$\frac1{2\sin\frac\pi{14}}$&$\frac{\sin\frac\pi7}{\cos\frac{3\pi}{14}}$&$-\frac1{2\sin\frac{3\pi}{14}}$&$-\frac1{2\sin\frac\pi{14}}$&$-\frac1{2\sin\frac\pi{14}}$&$-\frac{\sin\frac\pi7}{\cos\frac{3\pi}{14}}$&$-\frac{\sin\frac\pi7}{\cos\frac{3\pi}{14}}$&$\frac1{2\sin\frac{3\pi}{14}}$&$\frac1{2\sin\frac{3\pi}{14}}$&$-\frac1{2\sin\frac\pi{14}}$&$-\frac1{2\sin\frac\pi{14}}$&$-\frac{\sin\frac\pi7}{\cos\frac{3\pi}{14}}$&$-\frac{\sin\frac\pi7}{\cos\frac{3\pi}{14}}$&$\frac1{2\sin\frac{3\pi}{14}}$&$\frac1{2\sin\frac{3\pi}{14}}$\\
	$\hat{\mcal L}_{\frac34,\frac34}$&$1+\sqrt2$&$1+\sqrt2$&$1+\sqrt2$&$1-\sqrt2$&$1-\sqrt2$&$1-\sqrt2$&$1+\sqrt2$&$1+\sqrt2$&$1+\sqrt2$&$1-\sqrt2$&$1-\sqrt2$&$1-\sqrt2$&$1+\sqrt2$&$1+\sqrt2$&$1+\sqrt2$&$1+\sqrt2$&$1+\sqrt2$&$1+\sqrt2$&$1-\sqrt2$&$1-\sqrt2$&$1-\sqrt2$&$1-\sqrt2$&$1-\sqrt2$&$1-\sqrt2$\\
	$\hat{\mcal C}$&$\sqrt{2+\sqrt2}$&$-\sqrt{2+\sqrt2}$&$\sqrt{2+\sqrt2}$&$2\sin\frac\pi8$&$-2\sin\frac\pi8$&$2\sin\frac\pi8$&$-2\sin\frac\pi8$&$2\sin\frac\pi8$&$-2\sin\frac\pi8$&$-\sqrt{2+\sqrt2}$&$\sqrt{2+\sqrt2}$&$-\sqrt{2+\sqrt2}$&$\sqrt{2+\sqrt2}$&$-\sqrt{2+\sqrt2}$&$-\sqrt{2+\sqrt2}$&$\sqrt{2+\sqrt2}$&$\sqrt{2+\sqrt2}$&$-\sqrt{2+\sqrt2}$&$2\sin\frac\pi8$&$-2\sin\frac\pi8$&$-2\sin\frac\pi8$&$2\sin\frac\pi8$&$2\sin\frac\pi8$&$-2\sin\frac\pi8$
\end{tabular}.}
\end{table}
\end{landscape}
The relevant spin contents are given by
\begin{align*}
    \mcal H_F:&\quad s\in\{0\}\text{ mod }1,\\
    \mcal H_{\mcal L_{\frac5{14},\frac5{14}}}:&\quad s\in\{0,\pm\frac1{14},\pm\frac17,\pm\frac5{14},\pm\frac37,\pm\frac12\}\text{ mod }1,\\
    \mcal H_{\mcal L_{\frac97,\frac97}}:&\quad s\in\{0,\pm\frac27,\pm\frac37\}\text{ mod }1,\\
    \mcal H_{\mcal L_{\frac{34}7,\frac{34}7}}:&\quad s\in\{0,\pm\frac17,\pm\frac37\}\text{ mod }1,\\
    \mcal H_{\mcal L_{\frac{39}{14},\frac{39}{14}}}:&\quad s\in\{0,\pm\frac1{14},\pm\frac3{14},\pm\frac27,\pm\frac37,\pm\frac12\}\text{ mod }1,\\
    \mcal H_{\mcal C}:&\quad s\in\{-\frac{15}{32},-\frac{11}{32},-\frac3{32},\frac1{32},\frac5{32},\frac{13}{32}\}\text{ mod }1.
\end{align*}

\section{$\mbb Z_N$ paraspin modular category}\label{PMC}
In this appendix, we define $\mbb Z_N$ paraspin modular category, and study its properties. Almost all discussions proceed as in $\mbb Z_2$ case.

For $N>2$, we define a $\mbb Z_N$ parafermion in a unitary ribbon fusion category (URFC) as a primitive $\mbb Z_N$ simple object with $\mbb Z_N$ topological twist. Namely, a simple object $p$ with $p^N=1$ and $\theta_p=e^{\frac{2\pi in}N}$, with $\gcd(N,n)=1$. We call a unitary modular category (UMC) $\mcal C$ a $\mbb Z_N$ paraspin modular category (PMC) if $\mcal C$ has a $\mbb Z_N$ parafermion $p$, and denote it by a pair $(\mcal C,p)$.

The simplest non-trivial example is the case $N=3$. The critical and tricritical three-state Potts models are described by $\mbb Z_3$ PMCs $(\mcal C,\eta)$.

Following the Proposition 2.10 of \cite{fMTC}, we can prove some properties. Before we present the statements and their proofs, we first review actions of lines. We define an action of parafermion $p$ on a simple object $j$ by
\begin{equation}
    \ep^p_j:=\frac{\left(\widetilde S_\text{top}\right)_{p,j}}{\left(\widetilde S_\text{top}\right)_{1,j}}.\label{epp}
\end{equation}
One can visualize the quantity as follows. We can rewrite the definition as $\left(\widetilde S_\text{top}\right)_{p,j}=\ep^p_j\left(\widetilde S_\text{top}\right)_{1,j}=\ep^p_jd_j$. The LHS is nothing but the expectation value of linked knots of $p$ and $j$. Comparing this configuration against another without the $p$-loop, i.e., $\left(\widetilde S_\text{top}\right)_{1,j}$, we can interpret $\ep^p_j$ as the action of $p$-loop on $j$-loop. (One can also view the expectation value of $j$-loop in the denominator is inserted to cancel the quantum dimension of $j$.) One can show $\ep^p_j$ is a $\mbb Z_N$ phase.\footnote{Mathematically, this would be proved employing functoriality of braidings. Here, we instead give an intuitive proof. The $N$-th power of the monodromy charge $\left(\ep^p_j\right)^N$ is given by a $j$-loop with $N$ $p$-loops linked to it. There are two ways to evaluate this quantity, either one acts $N$ $p$-loops on $j$-loop one after another, or first fuses $N$ $p$-loops and acts the result on the $j$-loop. The second method gives $1$ because $p^N=1$. Thus the first method $\left(\ep^p_j\right)^N$ should also give $1$, $\left(\ep^p_j\right)^N=1$.} The phase is called monodromy charge \cite{SY90,FRS04,fMTC} or fusion character \cite{K05}. With this definition, we can prove the following:

$Proposition$. Let $p$ be a $\mbb Z_N$ parafermion in URFC $\mcal B$, then for any $j,k\in\mcal B$,\\
(i) $\left(\widetilde S_\text{top}\right)_{p,j}=\ep^p_jd_j$, (equivalently $c_{j,p}c_{p,j}=\ep^p_j
id_{pj}$),\\
(ii) $\theta_{pj}=\ep^p_j\theta_p\theta_j$,\\
(iii) $\left(\widetilde S_\text{top}\right)_{pj,k}=\ep^p_k\left(\widetilde S_\text{top}\right)_{j,k}$,\\
(iv) $\ep^p_{pj}=\ep^p_p\ep^p_j$.
\newline

$Proof$. (i) Obvious from the definition of the monodromy charge.\\
(ii) Evaluate the monodromy charge using the double braiding:
\begin{align*}
    \ep^p_j&\equiv\frac{\left(\widetilde S_\text{top}\right)_{p,j}}{\left(\widetilde S_\text{top}\right)_{1,j}}\\
    &=\frac1{d_j}\tr(c_{j,p}c_{p,j})\\
    &=\frac1{d_j}\frac{\theta_{pj}}{\theta_p\theta_j}d_{pj}=\frac{\theta_{pj}}{\theta_p\theta_j}.
\end{align*}
In the last equality, we used $d_j=d_{pj}$.\\
(iii) By definition of the topological $S$-matrix, we have
\[ \left(\widetilde S_\text{top}\right)_{pj,k}\equiv\tr(c_{k,pj}c_{pj,k}). \]
This is an expectation value of two linked knots, $pj$-loop and $k$-loop. Here, we undo the fusion $pj$, and act $p$ on the $k$-loop. The action gives $\ep^p_k$ and the remaining two linked knots, $j$-loop and $k$-loop, give the topological $S$-matrix:
\[ \left(\widetilde S_\text{top}\right)_{pj,k}=\ep^p_k\left(\widetilde S_\text{top}\right)_{j,k}. \]
More precisely, this can be shown employing the functoriality of braiding.\\
(iv) By definition and (iii), we have
\begin{align*}
    \ep^p_{pj}&\equiv\frac{\left(\widetilde S_\text{top}\right)_{p,pj}}{\left(\widetilde S_\text{top}\right)_{1,pj}}\\
    &=\frac1{d_j}\ep^p_p\left(\widetilde S_\text{top}\right)_{j,p}\equiv\ep^p_p\ep^p_j.
\end{align*}$\square$\\

A reader may notice that we have skipped the first statement of the proposition 2.10 in \cite{fMTC}. In case of SMC with a fermion $f$, the action of $f$ on itself is completely determined due to $f^2=1$:
\begin{align*}
    \ep^f_f&=\frac{\left(\widetilde S_\text{top}\right)_{f,f}}{\left(\widetilde S_\text{top}\right)_{1,f}}\\
    &=\tr(c_{f,f}c_{f,f})\\
    &=\frac{\theta_1}{\theta_f^2}=1.
\end{align*}
On the other hand, in case of $\mbb Z_N$ PMC, the action $\ep^p_p$ depends on the topological twist (or equivalently spin) of $p^2$:
\begin{align*}
    \ep^p_p&=\frac{\left(\widetilde S_\text{top}\right)_{p,p}}{\left(\widetilde S_\text{top}\right)_{1,p}}\\
    &=\tr(c_{p,p}c_{p,p})\\
    &=\frac{\theta_{p^2}}{\theta_p^2}.
\end{align*}
More explicitly, if we write the topological twist $\theta_{p^2}=e^{\frac{2\pi in_2}N}$, the monodromy charge is given by
\begin{equation}
    \ep^p_p=e^{\frac{2\pi i(n_2-2n)}N}.\label{eppp}
\end{equation}
Therefore, depending on the spin of $p^2$, $n_2$ (and also $n$), we have richer grading structure than in SMC. We will call a $\mbb Z_N$ PMC faithful if $\ep^p_p$ is a primitive $N$-th root of unity. Unfaithful $\mbb Z_N$ PMC has further classifications depending on $N$. For example, when $N=6$, we can also have $\ep^p_p=1,-1,e^{\frac{2\pi i}3}$. We will call the case $\ep^p_p=1$ trivial. Denoting the ``charge'' (mod $N$) of an object $j$ under $p$ as $|j|$, $\ep^p_j=e^{\frac{2\pi i|j|}N}$, from the proposition (iv), we learn tensoring with $p$ adds charge $|p|$, i.e., for $j\in\mcal B_{|j|}$, $pj\in\mcal B_{|p|+|j|}$ where $\mcal B_m$ is a collection of objects with charge $m$ defined mod $N$.

(One can show similar statements for $\ep^{p^l}_j$ by replacing $p$ by its $l$-th power $p^l$.)

For concreteness, let us study the case $N=3$. This describes three-state Pottes models. When $N=3$, we can fix $\ep^p_p$. This is because the spin of $p^2$ is fixed by the axiom of ribbon structure: $\theta_{p^2}=\theta_{p^*}\equiv\theta_p$. Hence we have
\[ \ep^p_p=\frac{\theta_{p^2}}{\theta_p^2}=\theta_p^{-1}. \]
In other words, $\mbb Z_3$ PMCs are necessarily faithful. (An example like the $(E_6)_3$ WZW model is not paraspin.) The critical and tricritical three-state Potts models correspond to $n=2,1$ (or equivalently $\ep^p_p=e^{\frac{2\pi i}3},e^{\frac{4\pi i}3}$), respectively.


\begin{thebibliography}{30}
\bibitem{DG12}
  T.~Dimofte and D.~Gaiotto,
  ``An E7 Surprise,''
  JHEP \textbf{10}, 129 (2012)
  doi:10.1007/JHEP10(2012)129
  [arXiv:1209.1404 [hep-th]].
\bibitem{MS1}
  G.~W.~Moore and N.~Seiberg,
  ``Classical and Quantum Conformal Field Theory,'' Commun. Math. Phys. \textbf{123}, 177 (1989) doi:10.1007/BF01238857
\bibitem{MS2}
  G.~W.~Moore and N.~Seiberg,
  ``LECTURES ON RCFT,'' RU-89-32.
\bibitem{K21}
  K.~Kikuchi,
  ``Symmetry enhancement in RCFT,''
  [arXiv:2109.02672 [hep-th]].
\bibitem{GKSW14}
  D.~Gaiotto, A.~Kapustin, N.~Seiberg and B.~Willett,
  ``Generalized Global Symmetries,''
  JHEP \textbf{02}, 172 (2015)
  doi:10.1007/JHEP02(2015)172
  [arXiv:1412.5148 [hep-th]].
\bibitem{BT17}
  L.~Bhardwaj and Y.~Tachikawa,
  ``On finite symmetries and their gauging in two dimensions,''
  JHEP \textbf{03}, 189 (2018)
  doi:10.1007/JHEP03(2018)189
  [arXiv:1704.02330 [hep-th]].
\bibitem{CLSWY}
  C.~M.~Chang, Y.~H.~Lin, S.~H.~Shao, Y.~Wang and X.~Yin,
  ``Topological Defect Lines and Renormalization Group Flows in Two Dimensions,''
  JHEP \textbf{01}, 026 (2019) doi:10.1007/JHEP01(2019)026
  [arXiv:1802.04445 [hep-th]].
\bibitem{C86}
  J.~L.~Cardy,
  ``Operator Content of Two-Dimensional Conformally Invariant Theories,''
  Nucl. Phys. B \textbf{270}, 186-204 (1986) doi:10.1016/0550-3213(86)90552-3
\bibitem{T18}
  Y.~Tachikawa, ``Topological Phases and Relativistic QFTs,'' https://member.ipmu.jp/yuji.tachikawa/lectures/2018-cern-rikkyo/.
\bibitem{HNT}
  C.~T.~Hsieh, Y.~Nakayama and Y.~Tachikawa,
  ``Fermionic minimal models,''
  Phys. Rev. Lett. \textbf{126}, no.19, 195701 (2021)
  doi:10.1103/PhysRevLett.126.195701
  [arXiv:2002.12283 [cond-mat.str-el]].
\bibitem{NY17}
  T.~Numasawa and S.~Yamaguchi,
  ``Mixed global anomalies and boundary conformal field theories,''
  JHEP \textbf{11}, 202 (2018) doi:10.1007/JHEP11(2018)202
  [arXiv:1712.09361 [hep-th]].
\bibitem{KY19}
  K.~Kikuchi and Y.~Zhou,
  ``Two-dimensional Anomaly, Orbifolding, and Boundary States,''
  [arXiv:1908.02918 [hep-th]].
\bibitem{BDLLS20}
  J.~B.~Bae, Z.~Duan, K.~Lee, S.~Lee and M.~Sarkis,
  ``Fermionic rational conformal field theories and modular linear differential equations,''
  PTEP \textbf{2021}, no.8, 08B104 (2021) doi:10.1093/ptep/ptab033 [arXiv:2010.12392 [hep-th]].
\bibitem{BDLLS21}
  J.~B.~Bae, Z.~Duan, K.~Lee, S.~Lee and M.~Sarkis,
  ``Bootstrapping fermionic rational CFTs with three characters,''
  JHEP \textbf{01}, 089 (2022)
  doi:10.1007/JHEP01(2022)089
  [arXiv:2108.01647 [hep-th]].
\bibitem{KCXC}
  K.~Kikuchi, J.~Chen, F.~Xu and C.~M.~Chang,
  ``Emergent SUSY in Two Dimensions,'' [arXiv:2204.03247 [hep-th]].
\bibitem{G12}
  D.~Gaiotto,
  ``Domain Walls for Two-Dimensional Renormalization Group Flows,'' JHEP \textbf{12}, 103 (2012) doi:10.1007/JHEP12(2012)103
  [arXiv:1201.0767 [hep-th]].
\bibitem{ENO}
  P.~Etingof, D.~Nikshych, V.~Ostrik, ``On fusion categories,'' [arXiv:math/0203060 [math.QA]].
\bibitem{W89}
  E.~Witten,
  ``Quantum Field Theory and the Jones Polynomial,''
  Commun. Math. Phys. \textbf{121}, 351-399 (1989)
  doi:10.1007/BF01217730
\bibitem{M12}
  M.~M\"uger,
  ``Modular Categories,'' [arXiv:1201.6593 [math.CT]].
\bibitem{K05}
  A.~Kitaev,
  ``Anyons in an exactly solved model and beyond,''
  Annals Phys. \textbf{321}, no.1, 2-111 (2006) doi:10.1016/j.aop.2005.10.005 [arXiv:cond-mat/0506438 [cond-mat.mes-hall]].
\bibitem{DW89}
  R.~Dijkgraaf and E.~Witten,
  ``Topological Gauge Theories and Group Cohomology,''
  Commun. Math. Phys. \textbf{129}, 393 (1990) doi:10.1007/BF02096988
\bibitem{T17}
  Y.~Tachikawa,
  ``On gauging finite subgroups,''
  SciPost Phys. \textbf{8}, no.1, 015 (2020)
  doi:10.21468/SciPostPhys.8.1.015
  [arXiv:1712.09542 [hep-th]].
\bibitem{O17}
  T.~Okazaki,
  ``Implications of Conformal Symmetry in Quantum Mechanics,''
  Phys. Rev. D \textbf{96}, no.6, 066030 (2017)
  doi:10.1103/PhysRevD.96.066030
  [arXiv:1704.00286 [hep-th]].
\bibitem{WK73}
  K.~G.~Wilson and J.~B.~Kogut,
  ``The Renormalization group and the epsilon expansion,''
  Phys. Rept. \textbf{12}, 75-199 (1974)
  doi:10.1016/0370-1573(74)90023-4
\bibitem{YZ89}
  V.~P.~Yurov and A.~B.~Zamolodchikov,
  ``TRUNCATED CONFORMAL SPACE APPROACH TO SCALING LEE-YANG MODEL,''
  Int. J. Mod. Phys. A \textbf{5}, 3221-3246 (1990) doi:10.1142/S0217751X9000218X
\bibitem{LM91}
  M.~Lassig and G.~Mussardo,
 ``Hilbert space and structure constants of descendant fields in two-dimensional conformal theories,''
  Comput. Phys. Commun. \textbf{66}, 71-88 (1991) doi:10.1016/0010-4655(91)90009-A
\bibitem{fMTC}
  P.~Bruillard, C.~Galindo, T.~Hagge, S.~H.~Ng, J.~Y.~Plavnik, E.~C.~Rowell and Z.~Wang,
  ``Fermionic Modular Categories and the 16-fold Way,''
  J. Math. Phys. \textbf{58}, no.4, 041704 (2017) doi:10.1063/1.4982048 [arXiv:1603.09294 [math.QA]].
\bibitem{SY90}
  A.~N.~Schellekens and S.~Yankielowicz,
  ``Simple Currents, Modular Invariants and Fixed Points,''
  Int. J. Mod. Phys. A \textbf{5}, 2903-2952 (1990) doi:10.1142/S0217751X90001367
\bibitem{FRS04}
  J.~Fuchs, I.~Runkel and C.~Schweigert,
  ``TFT construction of RCFT correlators. 3. Simple currents,''
  Nucl. Phys. B \textbf{694}, 277-353 (2004) doi:10.1016/j.nuclphysb.2004.05.014 [arXiv:hep-th/0403157 [hep-th]].
\bibitem{S16}
  N.~Seiberg,
  ``Continuum QFT methods in 2+1d,'' \href{https://www.youtube.com/watch?v=_fS-ErfPHsY}{The 34th Jerusalem Winter School in Theoretical Physics on New Horizons in Quantum Matter}.
\bibitem{CCHLS}
  Y.~Choi, C.~Cordova, P.~S.~Hsin, H.~T.~Lam and S.~H.~Shao,
  ``Non-Invertible Duality Defects in 3+1 Dimensions,''
  [arXiv:2111.01139 [hep-th]].
\bibitem{KNY21}
  M.~Koide, Y.~Nagoya and S.~Yamaguchi,
  ``Non-invertible topological defects in 4-dimensional $\mathbb {Z}_2$ pure lattice gauge theory,''
  PTEP \textbf{2022}, no.1, 013B03 (2022)
  doi:10.1093/ptep/ptab145
  [arXiv:2109.05992 [hep-th]].
\bibitem{RV98}
  P.~Ruelle and O.~Verhoeven,
  ``Discrete symmetries of unitary minimal conformal theories,'' Nucl. Phys. B \textbf{535}, 650-680 (1998) doi:10.1016/S0550-3213(98)00639-7
  [arXiv:hep-th/9803129 [hep-th]].
\bibitem{CW20}
  M.~Cheng and D.~J.~Williamson,
  ``Relative anomaly in ( 1+1 )d rational conformal field theory,''
  Phys. Rev. Res. \textbf{2}, no.4, 043044 (2020) doi:10.1103/PhysRevResearch.2.043044 [arXiv:2002.02984 [cond-mat.str-el]].
\bibitem{RW20}
  I.~Runkel and G.~M.~T.~Watts,
  ``Fermionic CFTs and classifying algebras,''
  JHEP \textbf{06}, 025 (2020) doi:10.1007/JHEP06(2020)025 [arXiv:2001.05055 [hep-th]].
\bibitem{ENO08}
  P.~Etingof, D.~Nikshych, V.~Ostrik,
  ``Weakly group-theoretical and solvable fusion categories,'' [arXiv:0809.3031 [math.QA]].
\bibitem{TW21}
  R.~Thorngren and Y.~Wang,
  ``Fusion Category Symmetry II: Categoriosities at $c$ = 1 and Beyond,''
  [arXiv:2106.12577 [hep-th]].
\bibitem{GK94}
  D.~Gepner and A.~Kapustin,
  ``On the classification of fusion rings,''
  Phys. Lett. B \textbf{349}, 71-75 (1995) doi:10.1016/0370-2693(95)00172-H
  [arXiv:hep-th/9410089 [hep-th]].
\bibitem{rank4}
  E.~Rowell, R.~Stong, Z.~Wang,
  ``On classification of modular tensor categories,'' [ 	arXiv:0712.1377 [math.QA]].
\bibitem{rank5}
  P.~Bruillard, S.H.~Ng, E.C.~Rowell, and Z.~Wang, ``ON CLASSIFICATION OF MODULAR CATEGORIES BY RANK,'' [arXiv:1507.05139 [math.QA]].
\bibitem{YF20}
  Y.~Yao and A.~Furusaki,
  ``Parafermionization, bosonization, and critical parafermionic theories,''
  JHEP \textbf{04}, 285 (2021) doi:10.1007/JHEP04(2021)285
  [arXiv:2012.07529 [cond-mat.str-el]].
\bibitem{K20}
  J.~Kulp,
  ``Two More Fermionic Minimal Models,''
  JHEP \textbf{03}, 124 (2021)
  doi:10.1007/JHEP03(2021)124
  [arXiv:2003.04278 [hep-th]].
\bibitem{ALW17}
  D.~Aasen, E.~Lake and K.~Walker,
  ``Fermion condensation and super pivotal categories,''
  J. Math. Phys. \textbf{60}, no.12, 121901 (2019)
  doi:10.1063/1.5045669
  [arXiv:1709.01941 [cond-mat.str-el]].
\end{thebibliography}
\end{document}